\documentclass[journal]{IEEEtran}
\ifCLASSINFOpdf
\else
\fi
\usepackage{comment}
\usepackage[utf8]{inputenc}
\usepackage[T1]{fontenc}
\usepackage[cmex10]{amsmath} % Use the [cmex10] option to ensure complicance
\usepackage{subfigure}
\usepackage{amssymb}
\usepackage{bm}
\usepackage{float}
\usepackage{graphicx}
\usepackage[noadjust]{cite}
\usepackage{multirow}
\usepackage{enumerate}
\usepackage{epstopdf}
\usepackage{color}

\usepackage{xcolor}
\definecolor{g}{rgb}{0.1,0.6,0.3}
\definecolor{r1}{rgb}{0.8,0.1,0.1}
\definecolor{r2}{rgb}{1,0.4,0}
\definecolor{b}{rgb}{0.2,0.3,1}

\usepackage{diagbox}

\allowdisplaybreaks

\newtheorem{exmp}{Example}
\newtheorem{theorem}{Theorem}
\newtheorem{lemma}[theorem]{Lemma}
\newtheorem{proposition}[theorem]{Proposition}

\newtheorem{remark}{Remark}

\newtheorem{definition}[theorem]{Definition}

\usepackage{comment}
\begin{document}

\title{Schedule Sequence Design for Broadcast in Multi-channel Ad Hoc Networks
}


\author{Fang~Liu,
        Kenneth~W.~Shum,~\IEEEmembership{Senior~member,~IEEE},
        Yijin~Zhang,~\IEEEmembership{Senior~member,~IEEE},
        and~Wing~Shing~Wong,~\IEEEmembership{Fellow,~IEEE}        
\thanks{The research was partially funded by Schneider Electric, Lenovo Group (China) Limited and the Hong Kong Innovation and Technology Fund (ITS/066/17FP) under the HKUST-MIT Research Alliance Consortium.}
\thanks{F. Liu and W. S. Wong are with the Department of Information Engineering, the Chinese University of Hong Kong, Hong Kong, China (e-mail: \{lf015, wswong\}@ie.cuhk.edu.hk).}
\thanks{K. W. Shum is with the School of Science and Engineering, the Chinese University of Hong Kong (Shenzhen), Shenzhen, China (e-mail: wkshum@cuhk.edu.cn).}
\thanks{Y. Zhang is with the  School of Electronic and Optical Engineering, Nanjing University of Science and Technology,
Nanjing, China (yijin.zhang@gmail.com).}
}


\maketitle

\begin{abstract}

We consider a single-hop ad hoc network in which each node aims to broadcast packets to its neighboring nodes by using multiple slotted, TDD collision channels. There is no cooperation among the nodes. To ensure successful broadcast, we propose to pre-assign each node a periodic sequence to schedule transmissions and receptions at each time slot. These sequences are referred to as schedule sequences. Since each node starts its transmission schedule independently, there exist relative time offsets among the schedule sequences they use. Our objective is to design schedule sequences such that each node can  transmit at least one packet to each of its neighbors successfully within a common period, no matter what the time offsets are. The sequence period should be designed as short as possible. In this paper, we analyze the lower bound on sequence period, and propose a sequence construction method by which the period can achieve the same order as the lower bound.

We also consider the random scheme in which each node transmits or receives on a channel at each time slot with a pre-determined probability. The frame length and broadcast completion time under different schemes are compared by numerical studies. 
\end{abstract}

\section{Introduction}
\subsection{Overall scenario}
We consider a medium access control (MAC) problem for a wireless
single-hop ad hoc network, in which each node always has a stream of packets to broadcast to its neighboring nodes. The nodes are within a common hearing range.
This broadcast scenario is common. For example, in a sensor network, each sensor node is required to collect data such as temperature and humidity observed by itself and other neighboring nodes for further processing \cite{wang2011reliable,cheng2016achieving,sherazi2018comprehensive}. Another example comes from vehicular ad hoc networks (VANETs), in which each vehicle broadcasts safety messages such as its speed and location information to its neighboring vehicles, in order to avoid collisions among the vehicles~\cite{zheng2015heterogeneous, al2014comprehensive,ali2018analysis}.

The overall objective of the MAC design is to ensure that any node can successfully receive broadcast packets from all other nodes within a short time duration. This is in line with the goal of ultra-reliable low latency communications (URLLC) in the fifth generation (5G) networks  \cite{bennis2018ultrareliable}. Lots of scheduling algorithms for single-channel environments have been proposed in the literature \cite{borgonovo2004adhoc,wu2013protocol,wu2014safety}. In this paper, we mainly focus on the multi-channel case. Specifically, we assume that the broadcast packets are transmitted over multiple slotted, time division duplex (TDD), equal-bandwidth collision channels. The TDD assumption indicates that each node at any time slot can either receive or transmit a packet over a channel but not both. 
A broadcast at a given channel is successfully received if during the whole transmission duration it is free from conflicts with other transmissions on the channel and the intended receiver is tuned to receive packets at the same channel. 
Compared with a single-channel system, the use of multiple channels has two main influences. On one hand, it enables the possibility of concurrent successful transmissions among multiple node pairs, and thus may expedite successful all-to-all broadcasts. On the other hand, 
for any node pair, the transmitter and the receiver should be matched to the same channel before data transmission. This matching process, referred to as \textit{rendezvous} \cite{zhang2019quaternary,cai2016strictly,sheu2018multi}, may result in longer delay, especially when their schedules are not under centralized control. 
Therefore, the problem that whether using more channels is beneficial for decreasing delay compared with using a single channel is nontrivial.

For broadcast in multi-channel ad hoc networks without centralized controller, most of the existing MAC schemes rely on  coordination among the nodes, which usually requires  control message exchange on a dedicated control channel \cite{klingler2016mcb,hadded2015tdma,zhang2014scalable}. 
 For example, MCB proposed in \cite{klingler2016mcb} follows a split phase approach, that is, each node periodically switches between the control channel and one of six  service channels. Before broadcasting data, each node should select a service channel for data transmission and should announce this information to other nodes through the control channel. However,  this control channel would be a bottleneck when traffic is heavy, and the overhead for frequent control message exchange would be high especially when the data packets are short packets \cite{maitra2016comparative}. 
 
In this paper, we aim at devising  multi-channel MAC schemes without centralized controller and negotiation among the nodes. 
For such a system, accurate time synchronization among the nodes is challenging to achieve.
Therefore, it is desirable to devise asynchronous MAC schemes. 
Without time synchronization, 
each node starts its transmission mechanism independently. It follows that the time difference between the start point of a node and the system-wide reference point $ t=0 $
may vary from node to node. We refer to this time difference as the \textit{time offset} of a node. The values of time offsets are unknown and remain unchanging during the whole communication session. 
% To facilitate discussions, we assume that the slot boundaries of the nodes are aligned, and thus the time offsets can only assume values that are multiples of a slot duration.

%This fully asynchronous case can be simplified to the case where slot boundaries of the nodes are aligned and thus time offsets can only assume values that are multiples of a slot duration. Previous studies \cite{chang2019asynchronous,massey1985collision}  have shown that the results obtained in this simplified case can be easily extended to the fully asynchronous case. Therefore, to facilitate discussions, we assume that the slot boundaries are aligned. Without loss of generality, we normalize the slot duration to 1 and assume all nodes start before the system time $t=0$, thus the time offsets can only be non-negative integers.

To the best of our knowledge, this is the first work that focuses on all-to-all broadcast in a multi-channel single-hop ad hoc network without synchronization and coordination.
We mainly consider deterministic schemes, and use random schemes as reference baseline. We  regard deterministic schemes as sequence schemes in which each node is pre-assigned a transmission and reception schedule in the format of a \textit{schedule sequence} \cite{shum2010construction,wu2014safety,liu2019sequence}. 
At each time slot, each node reads out its current sequence value, and then conducts corresponding action (transmitting or receiving on a particular channel) according to that value. An appropriately designed schedule sequence set can guarantee successful broadcasts within a common sequence period, for all possible time offsets. In the random schemes, each node at each time slot transmits or receives on a channel with a fixed probability. In both of the sequence schemes and random schemes, each node transmits or receives independently without cooperating with other nodes.

\subsection{Performance metrics}
The main goal in this paper is to devise MAC schemes to provide a hard guarantee on broadcast delay, for an asynchronous multi-channel network.
The metrics for broadcast delay are frame length and broadcast completion time, which are defined as follows. 
\subsubsection{Frame length}
\begin{figure}[htbp]
    \centering
    \includegraphics[width=3.2in]{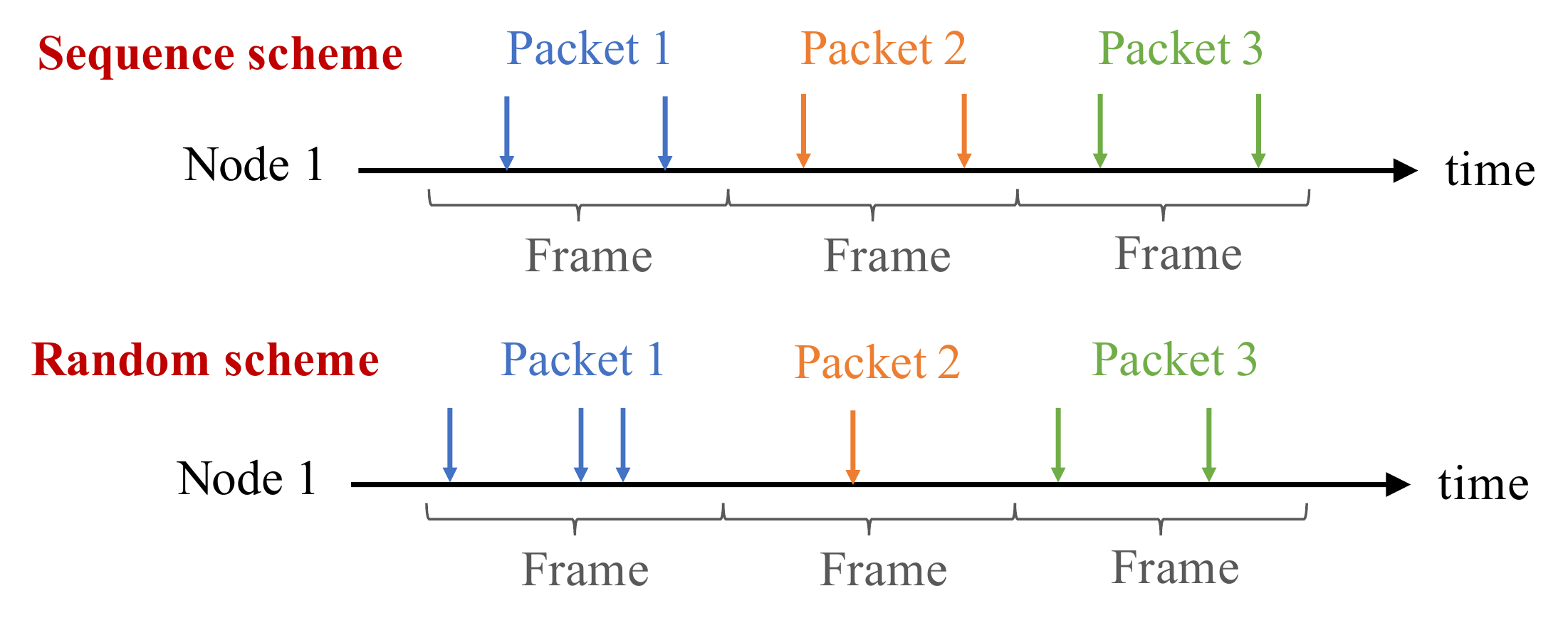}
  \caption{Each node transmits each packet in its packet stream for a frame duration. The arrows represent the time slots in which a packet is transmitted. In the sequence scheme, each packet is transmitted according to the periodic schedule sequence. 
  In the random scheme, each packet is transmitted random number of times within a frame.}
\label{fig:periodic traffic.}
\end{figure}

In the system we investigate, each node is required to transmit a sequence of packets to all other nodes. To ensure reliable communication, each node may need to transmit a packet for multiple times. We define the consecutive sequence of time slots in which the same packet is considered for transmission by a node as a frame, in both of the sequence schemes and the random schemes, as shown in Figure~\ref{fig:periodic traffic.}. 
%Each node can transmit a packet multiple times in a frame duration.
%The arrows in Figure~\ref{fig:periodic traffic.} represent the time slots in which a packet is transmitted. 

In a sequence scheme, each packet is transmitted according to a periodic schedule sequence. %In a random scheme, each packet is transmitted random number of times within a frame.
%Ideally, we aim to design sequences that can achieve one or more successful broadcast per frame for all nodes.
Our sequence design goal is to ensure that each node has one or more 
successful broadcasts per frame to each other node.
So for sequence-based schemes, the frame length is equal to the sequence period. 
For random schemes, the frame length represents the number of trials a 
node attempts to transmit a given packet, such that the probability 
that a successful broadcast can be achieved within a frame is close to 
1. (Here, the definition of closeness is determined according to
QoS requirements motivated by URLLC standards \cite{bennis2018ultrareliable}, since it it not 
possible to attain 100\% certainty for
random schemes.)
The frame length upper bounds the broadcast delay for all possible time offsets, so heuristically it should be minimized. 
%The design goal for both sequence schemes and random schemes is to minimize the frame length. 

\subsubsection{Broadcast completion time}
%We also compare sequence schemes and random schemes in terms of broadcast completion time, which is defined as the time duration starting from $ t=0 $ until each node has transmitted packets to each of its neighboring  nodes successfully. The comparison is mainly conducted 

The completion time is also a common metric for delay, which is usually referred to as \textit{group delay} for broadcast in the single channel model \cite{wu2013protocol,wu2014safety}. It is defined as the time duration starting from $ t=0 $ until each node has transmitted at least one packet to each other node successfully. The completion time varies with time offsets.
The comparisons for broadcast completion time under different schemes are mainly conducted by numerical studies. 

% With slot synchronization, the broadcast/unicast completion time varies with  time offsets. In the comparison, we generate time offsets randomly for a large number of times, and record the corresponding broadcast/unicast completion time. Then we compare the cumulative probability of broadcast/unicast completion time based on the records. 

\subsection{Related work}
\subsubsection{Broadcast with a single channel}
The design of schedule sequences for asynchronous broadcast was first studied for VANETs in \cite{wu2014safety}. 
%The proposed schedule sequences can enable each vehicle in a VANET to broadcast safety messages to its neighboring vehicles successfully at least once within a sequence period.
Various methods for assigning the sequences to vehicles have been discussed in \cite{wu2014safety, wong2013transmission, mao2019generalized}. %For example, sequence distribution can be conducted by the roadside node near highway entrances or toll booths. When a vehicle enters the highway or passes through a toll booth, the roadside node, which have stored a pool of schedule sequences in advance,  will assign the vehicle a unique sequence. 
Note that the schedule sequences proposed in \cite{wu2014safety} are only applicable for the single channel model and can be represented by binary protocol sequences with the User-Irrepressibility (UI) property, which have been extensively investigated in the literature (please see \cite{massey1985collision,gyorfi1993constructions,shum2010user,chen2018crt,chen2016constructions}  and references therein). The symbol value ``1'' or ``0'' in binary protocol sequences corresponds to transmitting or receiving on the single channel. The UI property signifies that for all possible time offsets, each protocol sequence has at least one~``1'' which does not collide with ``1''s from other sequences. Protocol sequences with the UI property can be constructed from conflict-avoiding codes \cite{jimbo2007conflict}. In this paper, we extend the analysis of protocol sequences to more general schedule sequences that are required for modeling multiple channel systems.  Using  multiple channels can increase throughput, however, sequence design also becomes more difficult due to the rendezvous process.

In \cite{wu2014safety}, the comparison between the proposed sequence scheme and the optimized random scheme in terms of broadcast completion time has been conducted. The comparison result is that the sequence scheme can achieve shorter broadcast completion time than the random scheme. 

\subsubsection{Unicast with multiple channels}
Sequence design for another common information exchange pattern, unicast, in asynchronous multi-channel system was investigated in  \cite{liu2019sequence}. The difference between unicast and broadcast lies in the  contents of the exchanged packets. In all-to-all broadcast, the packets transmitted from one node to other nodes are the same. As a contrast, in all-to-all unicast, the packets transmitted from one node to each of the other nodes are individual. That is, given $K$ nodes within the same hearing range,  then for the broadcast model, the total number of packets that should be transmitted successfully by the $K$ nodes within a frame is $K$, while for the unicast model, this value should be $K(K-1)$.

 The optimal transmitting and receiving probabilities for unicast under the random schemes are also analyzed in \cite{liu2019sequence}. The simulation results in \cite{liu2019sequence} show that under both sequence scheme and the optimized random scheme, the unicast completion time decreases when the number of available channels increases. Moreover, the unicast completion time under the sequence scheme is  shorter than that under the optimized random scheme. 
\subsection{Main contributions}
%In this paper, we consider MAC schemes for asynchronous all-to-all broadcast by multiple channels. 
%Therefore, how to utilize multi-channel resources efficiently to decrease broadcast delay is not  trivial. 
%Our major performance metric is frame length, which is defined as the time duration for a node to transmit a packet such that this packet can be successfully transmitted to all other nodes with high reliability. The frame length upper bounds the broadcast delay. Its value varies in different~schemes.

To the best of our knowledge, this is the first study that considers MAC schemes for asynchronous all-to-all broadcast by multiple channels without coordination among the nodes.
Our major contributions are listed as follows.
\begin{enumerate}
\item  
The following results are obtained for the sequence scheme. Given $K$ nodes and $M$ available channels,  we derive a lower bound on the shortest common sequence period, and propose a sequence design method based on the Chinese Remainder Theorem (CRT) correspondence. Under some general technical assumptions, the sequence period under our proposed construction has the same order as the lower bound, and can achieve an asymptotic reduction in the order of $M$ in comparison to the shortest known period for the single channel~case.
\item  We  analyze random schemes for benchmark against the sequence schemes.  We derive optimal transmitting and receiving probabilities for two random schemes.
\item Frame length and broadcast completion time under different schemes are compared by theoretical analysis and numerical studies. 

\end{enumerate}

%By comparison, we can observe that the sequence scheme can ensure broadcast in shorter delay while random schemes can only achieve broadcast in longer delay with a probability less than 1, when the two schemes are under same power consumption.

%We observe that the sequence scheme can ensure broadcast in shorter frame length than the random scheme.
%Moreover, we find that under the random scheme, the frame length with a single channel is shorter than that with multiple channels. Thus compared with the random scheme, our proposed sequence scheme can utilize the channel resources more efficiently.

%\item We also analyze random scheme as a comparison benchmark for the sequence scheme. By comparison, we can observe that the sequence scheme can ensure broadcast in shorter delay while random schemes can only achieve broadcast in longer delay with a probability less than 1, when the two schemes are under same power consumption. Moreover, we find that by the random scheme, the broadcast delay with a single channel is the shortest, which implies that the random scheme cannot utilize multi-channel resources efficiently.

The rest of this paper is organized as follows. After describing the system model in Section \ref{section: system model},  we present preliminary information in Section \ref{section: CRT sequences} to prepare for 
subsequent discussions. Then we analyze the lower bound on sequence period in Section~\ref{sec:B-lower bound}, and propose a  sequence construction method  in Section \ref{section: sequence design}. In Section \ref{section: comparison}, we present results on the sequence period by our proposed construction method under even group division. In Section~\ref{sec:random}, we analyze two random schemes. Comparisons on frame length and broadcast completion time under different schemes  are shown in Section~\ref{sec:B-compare}. Finally, we conclude the paper in Section~\ref{section: conclusion}.

%%%%%%%%%%%%%%%%%%%%%%%%%%%%%%%%%%%%%%%%%%%%%%%%%%%%%%%
\section{Problem Formulation} \label{section: system model}

We consider a single-hop ad hoc network consisting of $ K $ nodes that are all within a common hearing range. Each node should broadcast packets to each other node at least once within a sequence period, under the sequence scheme. 
%We note that this is the worst case for the defined problem in the sense that if a schedule design can guarantee successful broadcasts within a period $L$ for this case, then it can also provide such a guarantee if the nodes are distributed in a less dense manner over the region.
For notation simplicity in this paper, given a positive integer~$ n $, we use $[n]$ to denote the set $ \{1,2,\ldots,n \} $, and  $ \mathbb{Z}_{n} $ to denote the cyclic group $\{0,1,\ldots,n-1 \} $, with addition (resp. subtraction) modulo $n$ denoted by  $\oplus_n$ (resp.~$\ominus_n$).
We denote the $i$-th node by $N_{i}$, for $i \in [K]$. There are $ M $ frequency channels available. Since the bandwidth is a scarce resource in general, we only consider the case where $ M\leq K $ in this paper. 

We consider a group-based channel allocation method, called \textit{Assignment~T}. The~$ K $ nodes are divided into~$ M $ groups, denoted by $G_1$, $G_2, \ldots, G_M$. The group division satisfies $ \cup_{m=1}^{M}G_m=\{N_1,N_2,\ldots,N_K \} $, and $G_m\cap G_n = \emptyset$, for $m,n\in [M]$ and $m\neq n$.  
The group size of $G_m$ is denoted by $ |G_m| $.
We remark that the groups may have different sizes, and a group could be an empty set. 
If $G_m$ is empty, then channel $m$ would not be used.
Among the $M$ groups, we assume $G_1,G_2,\ldots, G_{W}$ are non-empty, $ 1\leq W \leq  M$, and denote the smallest (resp. largest) non-zero group size  by $ k $ (resp. $\ell$), i.e., 
\begin{align*}
k=\min \{ \vert G_1\vert, \vert G_2 \vert, \ldots, \vert G_{W} \vert \}, \\ \ell=\max \{ \vert G_1\vert, \vert G_2 \vert, \ldots, \vert G_{W} \vert \}. 
\end{align*}
The values of $W, k, \ell $ depend on how the groups are divided.
Especially, we define an even group division,  in which the division of the $W$ non-empty groups is as even as possible, that is, $ k= \left \lfloor K/W \right \rfloor$, $\ell=\left \lceil K/W \right \rceil$. 
Under a given group division, for $m=1,2,\ldots, W$, the nodes in group $G_m$ are allowed to transmit on channel~$m$ only, but are able to receive packets from any of the $W$ channels.% However, a node cannot transmit and receive simultaneously, i.e., it is either in a transmitting mode or receiving~mode at each time slot.

All time slots are assumed to be of equal duration. Without loss of generality, we normalize the slot duration to~1. We represent a periodic sequence with period~$ L $  by a sequence of finite length~$ L $. The schedule sequence of period $L$ assigned to node $ N_{i} $ is denoted~by 
$$\mathbf{s}_i:= [s_i (0) \ s_i(1) \ \ldots \ s_i(L-1)], $$
for $i=1,2,\ldots, K$. For node $N_i\in G_m$, where $ i \in [K]$ and  $ m\in [W]$, we denote the action to transmit on channel~$ m $ by the symbol $ T_{m} $, and the action to receive on channel~$ r $ by the symbol $ R_{r} $, for any $r\in [W]$, then the entries in $\mathbf{s}_i$ are chosen from the set
\begin{equation*}
 \{T_{m}\} \cup \{ R_{1}, R_{2},\ldots, R_{W} \}. 
\end{equation*}

For $i\in [K]$,  node $N_i$ has a time offset, denoted by $\tau_i$, which is defined as the time difference between the system-wide reference point $t=0$ and the starting point of  node $N_i$. 
% we assume that the slot boundaries of the $K$ nodes are aligned, and thus the time offsets can only assume values that are multiples of a slot duration.
To facilitate discussions, we assume that all nodes start their schedules no later than $t=0$ and that the slot boundaries of the nodes are aligned. As a result, the time offsets of the nodes are non-negative integers. Considering that the sequences have a common period~$ L $, we  assume that $ \tau_{i}\in \mathbb{Z}_{L} $. 
We let $ \bm{\tau} =(\tau_1,\tau_2,\ldots, \tau_K) \in \mathbb{Z}_{L}^{K} $ denote an instance of time offsets of the $K$ nodes.
For $N_i \in G_m$ with  $\tau_i$, we denote the cyclic shift of $\mathbf{s}_i$ by $\tau_i$ by
$$ \mathbf{s}_i^{\tau_i}:=[s_i(\tau_i) \ s_i(1\oplus_L \tau_i) \ \ldots \ s_i((L-1)\oplus_L \tau_i)].$$
If $s_i(t\oplus_L \tau_i)= T_m$,  node $N_i$ sends out a packet on channel $m$ at the time slot~$t$. If $s_i(t \oplus_L \tau_i)=R_r$, $ r\in [W] $, node $N_i$ listens to channel $r$ in time slot $t$ and see if any packet can be received. 
% If more than one node in $G_m$ reads a $ T_{m} $ from their sequences at a time slot, which implies that multiple nodes transmit on channel $ m $ simultaneously, then a collision occurs and no packets transmitted on channel~$ m $ at this time slot can be successfully decoded. 
If multiple nodes transmit on the same channel simultaneously, then a collision occurs and no packets transmitted on this channel at this time slot can be successfully decoded. 
For $m\in [W]$, if there is only one node transmitting on channel~$m$ and multiple nodes are receiving from channel~$m$ in the same time slot, the transmitted packet is regarded as successfully received by all the nodes that are listening to channel~$m$. 
%The packet is said to be broadcast to the receiving nodes.

%we denote the action to transmit on channel~$ m $ by~$ T_{m} $, and the action to receive on channel~$ r $ by~$ R_{r} $, for any $r\in \mathbb{I}_{M}$.
%Each entry in $ \mathbf{s}_{i}$ takes value from the set 
%The time offset of node~$ N_{i} $, denoted by~$ \tau_{i} $, can be any non-negative integer. 
%The value of $ \tau_{i} $ is unknown.
% The system is assumed to be slot-synchronous, thus $ \tau_{i} $ can be any non-negative integer. 
%Considering that the sequences have a common period~$ L $, we can assume that $ \tau_{i}\in \mathbb{Z}_{L} $. The combination of $ \tau_{i} $'s is represented by $\bm{\tau}=[\tau_{1}~\tau_{2}~\ldots~\tau_{K}]$, $\bm{\tau} \in \mathbb{Z}_{L}^{K}$. 

%In this case, node $ N_{i} $ transmits on channel $ m $ at slot $ t $ if $ s_{i}(t\oplus_{L}\tau_{i})=T_{m} $, and receives on channel $r$ at slot~$ t $ if $ s_{i}(t\oplus_{L}\tau_{i})=R_{r} $. If more than one node in $G_m$ reads a $ T_{m} $ from their sequences at a time slot, which implies that multiple nodes transmit on channel $ m $ simultaneously, then a collision occurs and no packets transmitted on channel~$ m $ at this time slot can be successfully decoded.

 %To facilitate discussions, we assume that $ K $ is a multiple of $ M $, thus each group contains exactly $ k=K/M $ nodes. 

The sequence design should ensure successful transmission between any two nodes within a period, regardless of the time~offsets. Specifically, the design of schedule sequences is subject to the following requirements.

\begin{enumerate} \item
 (Intra-group communication) For any $m\in [W]$, if $G_m$ has size $|G_m|\geq 2$, then for any $N_i, N_j \in G_m$ with $i\neq j$ and for any $\bm{\tau}\in\mathbb{Z}_L^K$, there exists a time index $t\in\mathbb{Z}_L$ such that
\begin{equation} \label{eq:B-requirement1}
\begin{cases}
\begin{split}
s_{i}(t\oplus_{L}\tau_{i})&=T_{m}; \\
s_{j}(t\oplus_{L}\tau_{j})&=R_{m}; \\
s_{x}(t\oplus_{L}\tau_{x})&\neq T_{m}, \text{ for all } N_x \in G_m\setminus\{N_i,N_j\}.
\end{split}
\end{cases}
\end{equation}
\item (Inter-group communication) For any two distinct group indices $m,n \in [W]$,  for any $N_i\in G_m$ and $N_j\in G_n$, and for any $\bm{\tau}\in\mathbb{Z}_L^K$, there exists a time index $t \in \mathbb{Z}_L$ such that
\begin{equation} \label{eq:B-requirement2}
\begin{cases}
\begin{split}
s_{i}(t\oplus_{L}\tau_{i})&=T_{m}; \\
s_{j}(t\oplus_{L}\tau_{j})&=R_{m}; \\
s_{x}(t\oplus_{L}\tau_{x})&\neq T_{m}, \text{ for all } N_x \in G_m \setminus\{ N_i\}.
\end{split}
\end{cases}
\end{equation}
\end{enumerate}

%The condition in \eqref{eq:B-requirement1} guarantees that any two nodes in the same group can exchange packets within a period,  regardless of the time offsets. 

%For given $ M $ channels and $ K $ nodes, to ensure that any node can  transmit at least one packet to each of its $ (K-1) $ neighboring nodes successfully within a period, $ L $, under Assignment T, we should design a sequence set $ \{\mathbf{s}_i: i\in [K]\}$  to satisfy the following criteria:
%for any transmitter-receiver pair, say transmitter node $N_i$ and receiver node $N_j$, where $N_i \in G_m$,  $i,j\in[K]$, $i\neq j$, $ m\in \mathbb{I}_{M} $, there exists a $ t\in \mathbb{Z}_{L} $ such that at this time slot, node~$N_i$ is transmitting on channel~$m$, node~$N_j$ is receiving on channel~$m$, and other nodes in~$G_m$ are not transmitting on channel $m$ to cause collisions with $N_i$, for all possible $ \bm{\tau} \in \mathbb{Z}_{L}^{K} $. That is, 
%\begin{align} \label{eq: B-requirement}
%\begin{cases}
%\begin{split}
%s_{i}(t\oplus_{L}\tau_{i})&=T_{m}; \\
%s_{j}(t\oplus_{L}\tau_{j})&=R_{m}; \\
%s_{x}(t\oplus_{L}\tau_{x})&\neq T_{m}, \text{ for } N_x \in G_m, x \in [K], x\neq i.
%\end{split}
%\end{cases}
%\end{align}

Given $K$ nodes and $M$ channels, 
a set of sequences $ \{\mathbf{s}_i: i\in [K]\}$ of length~$L$  is called an $ (M,K,L)$-\textit{schedule sequence set}~if there exists a positive integer $W\leq M$ and a partition  of $\{N_1,N_2,\ldots,N_K\}$ into $W$ non-empty groups $G_1 , G_2 ,\ldots, G_W$, so that 

\smallskip
(i) For each $ i\in [K] $, if $N_i \in G_m$, $m\in [W]$, the entries of sequence $\mathbf{s}_i$ are drawn from $\{T_m\}\cup\{R_1,R_2,\ldots, R_W\}$; 

\smallskip
(ii) The conditions in \eqref{eq:B-requirement1} and \eqref{eq:B-requirement2} are satisfied.

\medskip

%Because we will only consider channel allocation according to Assignment~$T$, a schedule sequence set satisfying the above two requirements will be simply called an $(M,K,L)$-schedule sequence set.
%The sequence length $L$ is an upper bound on the broadcast delay.

%Given integers $M$ and $K$ with $M\leq K$, we let $\Omega(M,K)$ denote the smallest length~$L$ such that an $ (M,K,L)$-schedule sequence set exists. 
%

\begin{comment}
\begin{example} \label{example1}
For an ad hoc network with 2 nodes ($ K=2 $) and 2 channels ($ M=2 $) under Assignment T, we let $G_1=\{N_1\}$ and $G_2=\{N_2\}$. That is, the node $ N_{1} $  can only transmit on channel 1, and the node $ N_2 $  can only transmit on channel 2. Here is a $ (2,2,4)$-schedule sequence set of length $L=4$:
\begin{align*}
\mathbf{s}_{1} =[T_{1}~~ T_{1}~~ R_{2}~~ R_{2}];~
\mathbf{s}_{2} =[T_{2}~~ R_{1}~~ T_{2}~~ R_{1}].
\end{align*}

We can check that for all $\tau_{1}, \tau_{2} \in \mathbb{Z}_4$, successful broadcast between the two nodes can be guaranteed. For example, if $ \tau_{1}=2 $, $\tau_{2}=1$, the shifted sequences $\mathbf{s}_{1}^{\tau_{1}}$ and $\mathbf{s}_{2}^{\tau_{2}}$ are as follows:
\begin{align*}
\mathbf{s}_{1}^2 =[ R_{2}~~ R_{2}~~T_{1}~~ T_{1}];~
\mathbf{s}_{2}^1 =[ R_{1}~~ T_{2}~~ R_{1}~~T_{2}].
\end{align*}
We can find that node $N_{2}$ can transmit to node $N_{1}$ on channel 2 when $t=1$, and node $N_{1}$ can transmit to node $N_{2}$ on channel 1 when $t=2$.
\end{example}
\end{comment}

\begin{exmp} \label{example2}
For 3 nodes ($ K=3 $) and 2 channels ($ M=2 $) under Assignment~T, we let $W=2$ and  $G_1=\{N_1, N_2\}, G_2=\{N_3\}$. Then the entries in $\mathbf{s}_1, \mathbf{s}_2$ are drawn from $\{T_1, R_1, R_2\}$ and the entries in $\mathbf{s}_3$ are drawn from $\{T_2, R_1\}$.
%That is, the nodes $ N_1, N_2 $  can only transmit on channel 1, and the node $ N_3 $  can only transmit on channel 2. 
Here is a $ (2,3,12)$-schedule sequence set of length $L=12$:
\begin{align*}
\mathbf{s}_{1}&=[T_1~T_1~T_1~T_1~T_1~T_1~R_1~R_1~
R_1~R_2~R_2~R_2];\\
\mathbf{s}_{2} &=[T_1~R_1~T_1~R_2~T_1~R_1~T_1~R_2~
T_1~R_1~T_1~R_2];\\
\mathbf{s}_{3}& =[T_2~R_1~R_1~T_2~R_1~R_1~T_2~R_1~
R_1~T_2~R_1~R_1].
\end{align*}

We can check that for all $\bm{\tau}\in \mathbb{Z}_{12}^3$, the conditions in \eqref{eq:B-requirement1} and \eqref{eq:B-requirement2} are satisfied. For example, if $ \bm{\tau}=(3,7,10)$, the shifted sequences $\mathbf{s}_{1}^{3}, \mathbf{s}_{2}^{7},\mathbf{s}_{3}^{10}$ are as follows:
\begin{align*}
\mathbf{s}_{1}^{3}&=[T_1~T_1~T_1~R_1~R_1~
R_1~R_2~R_2~R_2~T_1~T_1~T_1];\\
\mathbf{s}_{2}^{7}&=[R_2~T_1~R_1~T_1~R_2~
T_1~R_1~T_1~R_2~T_1~R_1~T_1];\\
\mathbf{s}_{3}^{10}& =[R_1~R_1~T_2~R_1~R_1~
T_2~R_1~R_1~T_2~R_1~R_1~T_2].
\end{align*}
We take $N_1$ for instance in this case.  It transmits to $N_2$ successfully at $t=2$ since $s_{1}^{3}(2)=T_1$ and $s_{2}^{7}(2)=R_1$, and transmits to $N_3$ successfully at $t=0$ since $s_{1}^{3}(0)=T_1, s_{2}^{7}(0)\neq T_1$ and $s_{3}^{10}(0)=R_1$.
\end{exmp}

In later sections, we will analyze lower bound on $L$ and propose construction method for $(M,K,L)$-schedule sequence~set. To facilitate reading, we list the notation introduced in this section in Table~\ref{tab:notation}.

\begin{table}[h]
%\center
%\begin{flushleft}
\caption{Notation Table}
%Notation Table
%\end{flushleft}
\label{tab:notation}
\begin{tabular}{c|c}
\hline
Notation & Definition \\
\hline
 $ K $ & The total number of nodes  \\
 $ M $ & The total number of available channels \\
  $G_m$ & The $m$-th group, $m\in [M]$ \\
 $ W $ & The number of non-empty groups, $ 1\leq W\leq M $\\
 $k $ & The smallest non-empty group size  \\
 $\ell $ & The largest non-empty group size \\
 $ L $ & The period of a periodic sequence set \\
 %$\Omega(M,K) $ & The smallest $L$ such that an $ (M,K,L)$-schedule sequence set exists \\
 $ N_i$ & The $i$-th node, $i\in [K]$ \\
 $\tau_i$ & The time offset of node $N_i$, $\tau_i \in \mathbb{Z}_{L}$  \\
 $\bm{\tau}$ & The combination of $\tau_i$'s, $\bm{\tau} \in  \mathbb{Z}_{L}^{K}$ \\
 
% $k_1, k_2$ & The group size  $k_1=\lceil K/M \rceil $, $k_2=\lfloor K/M  \rfloor$ \\
 $ \mathbf{s}_i $ & The schedule sequence assigned to node $N_i$   \\
% $s_i(t)$ & The $t$-th entry of $ \mathbf{s}_i $, $t\in \mathbb{Z}_{L}$ \\
 $\mathbf{s}_i^{\tau_i}$ & The cyclic shift of $\mathbf{s}_i$ by $\tau_i$\\
% $s_i(t\oplus_L \tau_i)$ & The $t$-th entry of $\mathbf{s}_i^{\tau_i}$, $t\in \mathbb{Z}_{L}$ \\
\hline
\end{tabular}
\end{table}

\section{Preliminaries} \label{section: CRT sequences}
In this section, we introduce preliminary information and present results that will be used in following sections.
\subsection{Hamming cross-correlation}

We introduce the definition and a basic property of the Hamming cross-correlation of two binary sequences. %The analysis for Hamming cross-correlation will be used for deriving the lower bound on $\Omega(M,K) $.

\begin{definition}
For two binary sequences $ \mathbf{s}_{1}:=[s_1(0)~s_1(1)~\ldots ~s_1(L-1)] $ and $ \mathbf{s}_{2}:=[s_2(0)~s_2(1)~\ldots ~s_2(L-1)] $ with common period~$ L $ and relative time offset~$ \tau $, $ \tau\in \mathbb{Z}_{L}$, their Hamming cross-correlation function is defined by
\begin{equation*}
H_{1,2}(\tau)=\sum_{t=0}^{L-1}s_{1}(t)s_{2}(t\oplus_{L}\tau).
\end{equation*}
When $ \mathbf{s}_{1}=\mathbf{s}_{2} $, $ H_{1,2}(\tau) $ is called the Hamming auto-correlation of $ \mathbf{s}_{1} $.
\end{definition}

\begin{definition}
The Hamming weight of a periodic binary sequence is defined as the number of ``1''s in a period.
\end{definition}

%For a periodic binary sequence, we use Hamming weight to denote the number of 1s in a period.
 Lemma \ref{lemma:pre-1} below illustrates a relationship between the Hamming cross-correlation and the Hamming weights of two binary sequences.

\begin{lemma} \label{lemma:pre-1}
\cite{sarwate1980crosscorrelation} For two binary sequences $ \mathbf{s}_{1} $, $\mathbf{s}_{2}$ with common period $L$ and with Hamming weights $w_1$, $w_2$, respectively, the sum of their Hamming cross-correlation, taken over relative time offset $ \tau $ ranging from 0 to $L-1$, satisfies
$$
\sum_{\tau=0}^{L-1}H_{1,2}(\tau)=w_1w_2.
$$
\end{lemma}

\subsection{CRT correspondence}\label{subsec: CRT}

We  remind readers of the Chinese Remainder Theorem (CRT) correspondence, since our proposed construction method for $ (M,K,L) $-schedule sequence set in Section~\ref{section: sequence design} is based on it.
\begin{definition}
For $ p $ and $ q $ that are relatively prime, the CRT correspondence is a bijective mapping between $ \mathbb{Z}_{pq} $ and  $ \mathbb{Z}_{p} \times \mathbb{Z}_{q}$ defined by
\begin{equation} \label{eq: CRT mapping}
\Phi_{p,q}(t):=(t \hspace{0.1cm}\text{mod} \hspace{0.1cm} p, t \hspace{0.1cm}\text{mod}\hspace{0.1cm} q).
\end{equation}
\end{definition}

By the CRT correspondence, a sequence of length $ L=pq $ can be obtained from a $ p\times q $ array with the $ (t\text{ mod }p, t\text{ mod }q) $-th entry in the array being mapped to the $ t $-th entry in the sequence, for $ t\in \mathbb{Z}_{L} $. Cyclically shifting the  sequence by $ \tau $, where $ \tau\in \mathbb{Z}_{L}$, is equivalent to row-wise and column-wise shifting its array representation by $ \tau \text{ mod } p $ and $ \tau \text{ mod } q $, respectively.

%\begin{comment}

%\end{comment}

\subsection{User-Irrepressible sequences}
User-Irrepressible (UI) sequences can be directly employed for broadcast in the single-channel model. They will also be used in our proposed construction method for $ (M,K,L)$-schedule sequence~set.

\begin{definition} \label{definition:UI sequences}
\cite{shum2010user} Consider a set of $K$ binary sequences each of which is of length $L$. We cyclically shift the $i$-th sequence by a time offset $ \tau_i\in \mathbb{Z}_L $, for $i\in [K]$, and stack these shifted sequences into a $ K\times L $ matrix $ \mathbf{M} $. If $\mathbf{M}$ always contains a $K\times K$ permutation matrix for all possible $\tau_1,\tau_2,\ldots,\tau_K \in \mathbb{Z}_L$, then this sequence set is a $(K,L)$-UI sequence set.
\end{definition}

%The reason why the UI sequence set can be used to schedule broadcast for single-channel systems is as follows. By Definition \ref{definition:UI sequences}, a $ (K,L) $-UI  sequence set has the property that for all possible $\tau_1,\tau_2,\ldots,\tau_K \in \mathbb{Z}_L$, each sequence has at least one slot in which its value is ``1'' and the other $(K-1)$ sequences are ``0'' within a period~$L$. If we assign a $ (K,L) $-UI sequence set to $K$ nodes, and let ``1'' represent transmitting on the channel and ``0'' represent receiving on the channel, then the UI sequence set can guarantee that each node has at least one slot in which it is transmitting on the channel and the other $(K-1)$ nodes are receiving on the channel within a period~$L$, which means that each node can successfully broadcast to the other $(K-1)$ nodes at least once. Thus a $ (K,L) $-UI sequence set is equivalent with a $ (1,K,L) $-schedule sequence set.

By Definition \ref{definition:UI sequences}, a $ (K,L) $-UI sequence set is equivalent with a $ (1,K,L) $-schedule sequence set. There are a variety of construction methods for UI sequence sets in the literature \cite{shum2010user,wu2014safety,lo2016partially}. %It turns out that UI sequences can be regarded as a transformation of conflict-avoiding codes  with a different perspective  \cite{momihara2007constant,jimbo2007conflict}.
%It turns out that UI sequences can be regarded under the perspective of conflict-avoiding codes \cite{momihara2007constant,jimbo2007conflict,mishima2009optimal,lin2016optimal,feng2019optimal}.
% UI sequences are also closely related to another combinatorial structure called cover-free family \cite{erdos1985families,furedi1996onr,ruszinko1994upper}.
It is well known that for a set of $K$ binary sequences  of length $L$, if the Hamming weight of each sequence is no less than $K$, and the Hamming cross-correlation between any two of them is no more than 1 for any time offsets, that is,
\begin{equation} \label{eq:UI-Hamming-correlation}
H_{i,j}(\tau_i\ominus_L \tau_j) \leq 1, \text{ for } i,j \in [K], i\neq j, \tau_i, \tau_j \in \mathbb{Z}_L,
\end{equation}
then this sequence set is a $ (K,L) $-UI sequence set.

For any given $K$, we can obtain a $(K,L)$-UI sequence set that satisfies \eqref{eq:UI-Hamming-correlation} by  the following construction, which is based on the CRT correspondence \eqref{eq: CRT mapping}.

\begin{definition}
\textbf{CRT-UI construction} \cite{shum2010user}: Given $ K $, let $ w\geq K $, $ p $ be a prime and $ p\geq w $, $ q $ be a number coprime with $ p $ and $ q\geq 2w-1 $. For generators $  g\in [K] $,   construct a set of $ K $  sequences $ \{ \mathbf{s}_{g}=[s_g(0)~\ldots ~s_g(L-1)]: g\in [K] \}$ with common Hamming weight $w$ and common period $ L=pq $ as follows: for $t\in \mathbb{Z}_L$,
\begin{equation} \label{eq:CRT construction}
s_{g}(t)=\begin{cases}
1 &\text{if } \Phi_{p,q}(t) = (ug \text{ mod }p, u \text{ mod }q), \text{ for }u\in \mathbb{Z}_{w}, \\
0 &\text{otherwise}.
\end{cases}
\end{equation}
\end{definition}

%By the CRT-UI construction, given $K$, when $K$ is a prime, the shortest period $ L $ of a $ (K,L) $-UI sequence set ($ (1,K,L) $-schedule sequence set) is
For any prime $K$, % the $ (1,K,L) $-schedule sequence set ($ (K,L) $-UI sequence set) obtained by the CRT-UI construction has a period that is no less than 
the shortest period $ L $ of a $ (1,K,L) $-schedule sequence set ($ (K,L) $-UI sequence set) obtained by the CRT-UI construction  is
\begin{equation} \label{eq:single channel}
L=K(2K-1).
\end{equation}
The period $L$ in \eqref{eq:single channel} is obtained by letting $w=K$, $p=K$ and $q=2K-1$. For general $K$ which may not be a prime, we can obtain the following equation on this shortest $ L $ by Bertrand's postulate,
\begin{equation} \label{eq:single channel-e1}
L\leq 2K(2K-1).
\end{equation}

The sequences obtained by the CRT-UI construction have the following Hamming auto-correlation property:
\begin{lemma} \label{auto-correlation}
\cite{shum2010construction} For $ g\in [p-1] $,  $ d\in \mathbb{Z}_{w} $ and $ \tau\in \mathbb{Z}_{L}$,
\begin{equation*}
H_{g,g}(\tau)=\begin{cases}
w-d &\text{if } \Phi_{p,q}(\tau)=\pm (g,1)d , \\
0 &\text{otherwise}.
\end{cases}
\end{equation*}
\end{lemma}

\begin{exmp} \label{example 1}
Given $ K=3 $, we design three sequences by the CRT-UI construction with  $ w=3 $, $ p=3 $, $ q=5 $, generators $g= 1,2,3 $ and length $ L=pq=15 $ as follows,
\begin{align*}
&\mathbf{s}_{1}=[1\ 1\ 1\ 0\ 0\ 0\ 0\ 0\ 0\ 0\ 0\ 0\ 0\ 0\ 0]; \\
&\mathbf{s}_{2}=[1\ 0\ 0\ 0\ 0\ 0\ 0\ 1\ 0\ 0\ 0\ 1\ 0\ 0\ 0]; \\
&\mathbf{s}_{3}=[1\ 0\ 0\ 0\ 0\ 0\ 1\ 0\ 0\ 0\ 0\ 0\ 1\ 0\ 0].
\end{align*}
Since the CRT-UI construction is based on the CRT correspondence, these sequences can be obtained from the following three $3\times 5 $ arrays, respectively:
\begin{equation} \label{array form}
\mathbf{s}_{1}: \setlength\arraycolsep{2.5pt}\begin{array}{|c|c|c|c|c|}
\hline
1 & 0 & 0 & 0 & 0 \\
\hline
0 & 1 & 0 & 0 & 0 \\
\hline
0 & 0 & 1 & 0 & 0 \\
\hline
\end{array}, ~
\mathbf{s}_{2}: \begin{array}{|c|c|c|c|c|}
\hline
1 & 0 & 0 & 0 & 0 \\
\hline
0 & 0 & 1 & 0 & 0 \\
\hline
0 & 1 & 0 & 0 & 0 \\
\hline
\end{array}, ~
\mathbf{s}_{3}: \begin{array}{|c|c|c|c|c|}
\hline
1 & 1 & 1 & 0 & 0 \\
\hline
0 & 0 & 0 & 0 & 0 \\
\hline
0 & 0 & 0 & 0 & 0 \\
\hline
\end{array}.
\end{equation}
%We take the sequence $ s_{2}(t) $ for instance. The values of $ t\in \mathbb{Z}_{15} $ that satisfy $ \Phi_{p,q}(t)=(2u \text{ mod }p,  u \text{ mod }q) $ corresponding to $ u=0,1,2 $ are $ t=0,11,7 $, respectively, thus $ 1 $s appear in $ s_{2}(t) $ when $ t=0,11,7 $. In its array representation, $ 1 $s are located in the positions $ (0,0)$, $(2,1) $ and $ (1,2) $.
We take the sequence $ \mathbf{s}_{2} $ for instance. In its array representation, ``1''s are located in the positions $\{(2u \text{ mod }p,  u \text{ mod }q): u=0,1,2 \}=\{(0,0),(2,1),(1,2) \}$. Since $ \Phi_{p,q}(0)=(0,0) $, $\Phi_{p,q}(11)=(2,1)$, $\Phi_{p,q}(7)=(1,2)$, thus ``1''s appear in $ \mathbf{s}_{2} $ when $ t=0,11,7 $.
%We denote the time offsets of the three sequences as $ \tau_{1} $, $ \tau_{2} $ and $ \tau_{3} $, respectively.

 If $ \tau_{2}=7 $, the shifted sequence $ \mathbf{s}_{2}^{7} $ is as follows,
\begin{equation*}
\mathbf{s}_{2}^{7}=[1\ 0\ 0\ 0\ 1\ 0\ 0\ 0\ 1\ 0 \ 0\ 0\ 0\ 0\ 0].
\end{equation*}
In $ \mathbf{s}_{2}^{7} $, ``1''s appear when $ t=8,4,0 $. Correspondingly, as shown in \eqref{eq:shifted s2}, ``1''s appear in the  positions $ (2,3), (1,4) $ and $ (0,0) $ in the array representation of $ \mathbf{s}_{2}^{7} $, which can be obtained by row-wise and column-wise shifting its original array representation in \eqref{array form} by $ (\tau_2 \text{ mod } p = 1 )$ and $ (\tau_2 \text{ mod } q=2) $, respectively.
\begin{equation} \label{eq:shifted s2}
\mathbf{s}_{2}^{7}: \setlength\arraycolsep{2.5pt}\begin{array}{|c|c|c|c|c|}
\hline
1 & 0 & 0 & 0 & 0 \\
\hline
0 & 0 & 0 & 0 & 1 \\
\hline
0 & 0 & 0 & 1 & 0 \\
\hline
\end{array}.
\end{equation}

We can check that, for any $ \tau_{1}, \tau_{2}, \tau_{3}\in \mathbb{Z}_{15} $, the Hamming cross-correlation of any two of $ \mathbf{s}_{1}^{\tau_1}$, $ \mathbf{s}_{2}^{\tau_2} $ and $ \mathbf{s}_{3}^{\tau_3} $ is $ 0 $ or $ 1 $. Thus $ \mathbf{s}_1$, $ \mathbf{s}_2 $ and $\mathbf{s}_3$ form a $ (3,15) $-UI sequence set.

We can also verify Lemma~\ref{auto-correlation}. We take the Hamming auto-correlation of $\mathbf{s}_2$ and $\mathbf{s}_2^7$ for instance. %When $\tau_1=2$,  $\Phi_{3,5}(\tau_1)=2\times(1,1)$, $ H_{1,1}(\tau_1)=1=w-2 $. This is in line with Lemma~\ref{auto-correlation}.
By definition, we have $ H_{2,2}(7)=1 $. On the other hand, $\Phi_{3,5}(7)=2\times(2,1)$, $ w-2=1 $. This is in line with Lemma~\ref{auto-correlation}.
\end{exmp}

\subsection{A result for a recursive sequence}
%We present a result for a recursive sequence. This will be used when we prove the lower bound on $L$ of an $ (M,K,L)$-schedule sequence~set in Theorem~\ref{theorem: B-lower bound for general M}.

In order to establish lower bound on sequence period $L$,
we need a technical lemma concerning a real valued sequence, which is
defined recursively as follows.
\begin{lemma}\label{thm: P-blocking}
Define a recursive sequence $ (b_{r})^{\infty}_{r=1} $ by
\begin{equation*}
   \begin{cases}
   b_{1}\geq C, \\
    b_{r}=b_{r-1}-\left \lceil \dfrac{b_{r-1} b_1 \mu}{L} \right\rceil, \text{ for } r \geq 2,
   \end{cases}
 \end{equation*}
where $C$ is a positive integer and $\mu$ is a real number that satisfies $ \mu \geq 1  $. If $ b_{C} \geq 1$,  then we have 
 \begin{equation*}
L \geq \left \lceil \dfrac{8C^{2}\mu}{9} \right\rceil .
 \end{equation*}
\end{lemma}

Please refer to Appendix~\ref{appendix 1} for the proof of Lemma~\ref{thm: P-blocking}.

%%%%%%%%%%%%%%%%%%%%%%%%%%%%%%%%%%%%%%%%%%%%%%%%%%%%%%%

\section{Lower Bound on Period $L$} \label{sec:B-lower bound}
%Given integers $M$ and $K$, for positive integer $W \leq M$, we define $\Omega(W,K,k) $ as the smallest length $L$ such that an $ (M,K,L)$-schedule sequence set exists for some partition $G_1,G_2,\ldots,G_W$ of $\{N_1,N_2,\ldots, N_K\}$. The analysis rely on~$k$, which is the smallest group size among the $W$ non-empty groups. 
Given a partition  of $\{N_1,N_2,\ldots,N_K\}$ into $W$ non-empty groups $G_1 , G_2 ,\ldots, G_W$
 with the smallest group size $k$, $ W\leq M $, we define $\Omega(W,k,M,K) $ as the smallest length $L$ such that an $ (M,K,L)$-schedule sequence set exists.

For a sequence set $\{\mathbf{s}_i:i\in [K]\}$, we denote the number of transmitting symbols in node $N_i$'s sequence $\mathbf{s}_i$ by $\alpha_i$, and the number of receiving symbols $R_r$'s in  $\mathbf{s}_i$ by $\beta_i^{r}$, for $i\in [K]$, $r\in [W]$. It is obvious that
\begin{equation} \label{eq:B-LB-e1}
L=\alpha_i+\sum_{r=1}^{W}\beta_i^{r}, \text{ for any } i\in [K]. 
\end{equation}
From \eqref{eq:B-LB-e1}, we can derive that there must exist an $i\in [K]$ and an $r\in [W]$ such that  
$$ \beta_i^{r}\leq L/W .$$
 Next we consider the nodes in group $G_r$, and denote the node with the smallest number of transmitting symbols $T_r$'s among nodes in $G_r$ by $N_{r_1}$.
% We denote these nodes by $ N_{r_1}, N_{r_2},\ldots,\\ N_{r_{|G_r|}} $, their assigned schedule sequences by $ \mathbf{s}_{r_1}, \mathbf{s}_{r_2},\ldots,\mathbf{s}_{r_{|G_r|}} $, and the number of transmitting symbols $T_r$'s in these sequences by $ \alpha_{r_1}, \alpha_{r_2},\ldots, \alpha_{r_{|G_r|}} $. Without loss of generality, we assume that $$\alpha_{r_1}=\min_{j\in[|G_r|]} \alpha_{r_j},$$ that is, node $N_{r_1}$  has the smallest number of transmitting symbols $T_r$'s among nodes in $G_r$. 
If this sequence set is an $ (M,K,L) $-schedule sequence set, then node $N_i$ can be guaranteed to receive a collision-free packet from node $N_{r_1}$ successfully within a period $L$. This means that at least one $T_r$ in $\mathbf{s}_{r_1}$ can match with an $R_r$ in $\mathbf{s}_i$, without colliding with $T_r$'s from other nodes in $G_r$.
% There are $(|G_r|-1)$ nodes in $G_r$ that would cause collisions on channel $r$. 
Considering that $|G_r|\geq k$ and that node $N_i$ may also in $G_r$, there should be at least $(k -2)$ competitors in $G_r$ besides $N_i$ and $N_{r_1}$ that would transmit on channel~$r$.  
%The more competitors that would cause collisions, the longer $L$ should be. 
%Thus we will analyze \textcolor{blue}{the case that} $N_i $ is also in $G_r$ and $|G_r|=k$.
% In this case, we use $N_{r_2}$ to denote~$N_i$. 
To facilitate subsequent discussions, we reduce these schedule sequences to binary sequences in the following way. For sequence $\mathbf{s}_{r_1}$, we replace the transmitting symbols $T_r$'s by ``1''s, and replace the other symbols by ``0''s. This newly obtained binary sequence is denoted by $\mathbf{e}_1$. For sequence $\mathbf{s}_{i}$, we replace all the non-$R_r$ symbols by ``1''s, and replace $R_r$'s by ``0''s. This newly obtained sequence is denoted by $\mathbf{e}_2$. For the $(k-2)$ sequences corresponding to the $(k-2)$ potential competitors, we replace $T_r$'s by ``1''s and other symbols by ``0''s. These $(k-2)$ newly obtained sequences
are denoted by $\mathbf{e}_3, \mathbf{e}_4, \ldots, \mathbf{e}_{k}$. The number of ``1''s in $\mathbf{e}_1$ is denoted by $a_1$, and the number of ``1''s in $\mathbf{e}_j$ is denoted by $ w_j $, for $j\in \{2,3,\ldots,k\}$. We have  
$$ w_2=L-\beta_i^{r}\geq L-L/W, ~ w_3, \ldots, w_{k} \geq a_1 .$$

In the following, we analyze the lower bound on $ \Omega(W,k,M,K)$ by using the blocking algorithm \cite{shum2010user}, in which we fix $\mathbf{e}_1$ and cyclically shift $ \mathbf{e}_2, \mathbf{e}_3, \ldots, \mathbf{e}_{k} $ to collide as many ``1''s in $\mathbf{e}_1$ as possible. 

\textbf{Blocking algorithm}

Inputs: A set of $k$ binary sequences $ \mathbf{e}_1, \mathbf{e}_2, \ldots, \mathbf{e}_{k} $ with common period~$L$. 

1. Set $j=2$.

2. Choose a time offset $\tau_j \in \mathbb{Z}_{L}$ for $ \mathbf{e}_{j} $ such that $ w_{j} $ ``1''s in $\mathbf{e}_j^{\tau_j}$ and $ a_{j-1}$ ``1''s in~$ \mathbf{e}_1 $ collide for the most number of times, that is, the Hamming cross-correlation between $\mathbf{e}_1$ and $ \mathbf{e}_j$ with relative time offset $\tau_j$, $H_{1,j}(\tau_j)$, is maximal.

3. Set the colliding ``1''s in $ \mathbf{e}_1 $ to ``0''s. Let $ a_{j} $ be the number of remaining ``1''s in $ \mathbf{e}_1 $ after colliding with $ \mathbf{e}_j $, $ a_j =a_{j-1}-\max\limits_{\tau_{j}\in \mathbb{Z}_{L}}H_{1,j}(\tau_{j})$.

4. If $j<k$, increase $j$ by one and go back to Step 2.

5. Output $ a_1, a_2,\ldots,a_{k} $ and stop.
\vspace{0.2cm}

By the blocking algorithm, the values of $ a_1, a_2,\ldots,a_{k} $ are non-negative.
If $ a_{k}=0 $, that is, none of the ``1''s in $\mathbf{e}_1$ can  match with a ``0'' in $\mathbf{e}_2^{ \tau_2}$ without colliding with ``1''s in $\mathbf{e}_j^{\tau_j}$'s, for $j\in \{3,4,\ldots,k\}$, then correspondingly none of the $T_{r}$'s in $
\mathbf{s}_{r_1}$  can match with an $R_r$ in $ \mathbf{s}_{i}^{\tau_{i}}$ without colliding with the $(k-2)$ potential competitors. Thus if the sequence set $\{\mathbf{s}_{i}:i\in [K]\}$ is an $ (M,K,L) $-schedule sequence set, then we must have $ a_{k}\geq 1$. Next we will analyze necessary condition for $ a_{k}\geq 1$.

There exists a relation among $ a_1, a_2,\ldots,a_{k} $ which is 
summarized by the following lemma.

\begin{lemma} \label{lemma: P-blocking}
By the blocking algorithm, we have 
\begin{equation} \label{eq: B-blocking1}
a_j\leq a_{j-1} -\left \lceil \dfrac{a_{j-1}w_j}{L} \right \rceil, \text{ for }j\in \{2,3,\ldots,k\}.
\end{equation}
\end{lemma}
\begin{IEEEproof}
By the blocking algorithm,  for $j\in \{2,3,\ldots,k\}$, $ a_j =a_{j-1}-\max\limits_{\tau_{j}\in \mathbb{Z}_{L}}H_{1,j}(\tau_{j})$.
%The sequence $(a_r)_{r=0}^{K-1}$ is monotonically non-increasing.
 By Lemma \ref{lemma:pre-1}, the sum of $ H_{1,j}(\tau_{j}) $ for all $ \tau_j \in \mathbb{Z}_{L} $ satisfies
$$
\sum_{\tau_{j}=0}^{L-1}H_{1,j}(\tau_{j})=a_{j-1} w_j.
$$
Thus
$$
\max\limits_{\tau_{j}\in \mathbb{Z}_{L}}H_{1,j}(\tau_{j}) \geq \left \lceil \dfrac{a_{j-1}w_j}{L} \right \rceil.
$$
This completes the proof for Lemma~\ref{lemma: P-blocking}. 
\end{IEEEproof}

Based on \eqref{eq: B-blocking1}, we define a recursive sequence $ (b_j)^{\infty}_{j=2} $ to make the analysis for $ a_{k}\geq 1$ more tractable. 

\begin{theorem}\label{theorem: B-blocking}
Define a sequence $ (b_{j})^{\infty}_{j=2} $ recursively by

\begin{equation*}
   \begin{cases}
   b_{2}= \left \lceil \dfrac{a_1}{W} \right \rceil, \\ 
   b_{j}=b_{j-1}-\left \lceil \dfrac{b_{j-1}b_2 W \varepsilon }{L} \right\rceil, \text{ for } j \geq 3,
   \end{cases}
 \end{equation*}
where $\varepsilon$ is a real number that satisfies $ b_2 W\varepsilon \leq a_1 $. Then, $a_j\leq b_j$, for $ j\in \{2,3,\ldots,k\}$.
\end{theorem}

\begin{IEEEproof}
We will prove $a_j\leq b_j$ for $ j\in \{2,3,\ldots,k\}$ by mathematical induction. At first, we consider the value of $a_2$. Due to \eqref{eq: B-blocking1} and the fact that $w_2\geq L-L/W$, we have 
\begin{equation}\label{eq: B-induction-e0}
a_2 \leq a_1 -\left \lceil \dfrac{a_1 (L-L/W)}{L} \right \rceil \leq a_1 -a_1 \left(1-\dfrac{1}{W}\right) =\dfrac{a_1}{W}. 
\end{equation}
Then we have $a_2\leq b_2$.

Next we assume $ a_{j-1} \leq b_{j-1}$ for $ j\in \{3,4,\ldots,k\} $. We have known that $ w_3, \ldots, w_{k} \geq  a_1$. Then by \eqref{eq: B-blocking1}, we have that for $ j\in \{3,4,\ldots,k\} $, 
\begin{equation} \label{eq: B-induction-e1}
a_j\leq a_{j-1} -\left \lceil \dfrac{a_{j-1}a_1}{L} \right \rceil.
\end{equation}
%Now consider a function $y=x-\lceil x \eta \rceil $, where $\eta \in [0,1)$ is a constant and  $x$ is a positive integer variable. This function is monotonically non-decreasing w.r.t $x$. 
%We can regard $a_1/L \in (0,1)$ as $\eta$. 
Since $a_1/L \in (0,1)$,  given $ a_{j-1} \leq b_{j-1}$, we have 
\begin{equation*} 
a_{j-1}-\left \lceil \dfrac{a_{j-1} a_1}{L} \right \rceil \leq b_{j-1}-\left \lceil \dfrac{b_{j-1} a_1}{L} \right \rceil .
\end{equation*}
Since $ b_2 W\varepsilon\leq a_1 $, we have
\begin{equation} \label{eq: B-induction-e2}
a_{j-1}-\left \lceil \dfrac{a_{j-1} a_1}{L} \right \rceil  \leq b_{j-1}-\left \lceil \dfrac{b_{j-1}b_2 W\varepsilon}{L} \right\rceil =b_j.
\end{equation}
By combining \eqref{eq: B-induction-e1} and \eqref{eq: B-induction-e2}, we  obtain that $ a_j\leq b_j $, for $ j\in \{3,4,\ldots,k\} $. This completes the proof.
\end{IEEEproof}

%Based on Theorem \ref{theorem: B-blocking}, if $ a_{k_2}\geq 1 $, then we must have $ b_{k_2}\geq 1 $. Next we obtain a lower bound on~$ L $  in Theorem \ref{theorem: B-lower bound for general M} by analyzing necessary condition for $b_{k_2}\geq 1$.

\begin{theorem} \label{theorem: B-lower bound for general M}
For an $ (M,K,L) $-schedule sequence set, we have
\begin{equation}  \label{eq: B-lower bound}
L \geq \left \lceil \dfrac{8 (k -1)^{2} W \varepsilon}{9} \right\rceil,
\end{equation}
where $ \varepsilon= 1-\dfrac{1}{k}$.
\end{theorem}

%Please see Appendix \ref{appendix B-2} for the proof of Theorem \ref{theorem: B-lower bound for general M}.

\begin{IEEEproof}
We first show that the value of $ \varepsilon $ can guarantee that 
\begin{equation} \label{eq: B-bound proof-e0}
 b_2 W\varepsilon\leq a_1 .
\end{equation}
Given $a_{k}\geq 1$, it follows that $ \max\limits_{\tau_{j}\in \mathbb{Z}_{L}}H_{1,j}(\tau_{j})\geq 1 $ for $ j\in \{3,4,\ldots,k\} $. Then we have $ a_2 \geq k -1 $.
%Then we must have $ a_2 \geq k_2 -1 $ since each of the other $(k_2 -2)$ sequences can collide at least one ``1'' in $ \mathbf{e}_1 $.
On the other hand, we have $a_2\leq a_1/W$ by \eqref{eq: B-induction-e0}. Therefore, we  obtain that $ a_1\geq W(k-1) $. 
Since $ \varepsilon=1-1/k $, we have 
$$
\dfrac{W\varepsilon}{1-\varepsilon}= W(k-1)\leq a_1,
$$
which by simple manipulation can be rewritten as
$$
\left( \dfrac{a_1}{W} +1 \right)W\varepsilon \leq a_1.
$$
Since $ b_2=\left \lceil a_1/W \right \rceil < a_1/W +1$, then \eqref{eq: B-bound proof-e0} can hold. 

With this $ \varepsilon $, we have $ b_{k}\geq a_{k}\geq 1$, as well as $ b_2\geq a_2\geq k-1 $
by Theorem~\ref{theorem: B-blocking}.  
Then  \eqref{eq: B-lower bound} follows by plugging $C=k -1$ and $\mu =W\varepsilon$ into Lemma~\ref{thm: P-blocking}.
\end{IEEEproof}

\begin{remark}
When the number of transmitting symbols in each sequence is a multiple of $W$, that is, $ \alpha_i $ is a multiple of $W$ for any $i\in [K]$, we then have $ b_2=\lceil a_1/W \rceil =a_1/W $. In this case, with $\varepsilon =1$, \eqref{eq: B-bound proof-e0} can be satisfied. Then the lower bound on $L$ can be improved to
\begin{equation*}
L \geq \left \lceil \dfrac{8 (k -1)^{2} W }{9} \right\rceil.
\end{equation*}
\end{remark}

We can observe from Theorem~\ref{theorem: B-lower bound for general M} that the lower bound \eqref{eq: B-lower bound} is loose when $k$ is small. Here we  provide another lower bound as a supplement.

\begin{theorem} \label{thm: B-bound2}
For an $ (M,K,L) $-schedule sequence set, if $ k\geq 2 $, we have
\begin{equation} \label{eq: B-bound2-e1}
L\geq 4W(k-1).
\end{equation}
If $k=1$, we have
$
L\geq 4(W-1).$
\end{theorem}

\begin{IEEEproof}
The least required period for $K$ nodes is no less than that for $K'=Wk$ nodes. We will analyze the lower bound on $L$ for $K'$ nodes.
Consider the transmission from node $N_i$ to node $N_j$, for $i,j\in [K']$, $i\neq j$.  Assume that $N_i\in G_m$, $m\in [W]$.
For the case of $k\geq 2$,
there are at least $(k -2)$ nodes in $G_m$ that would cause collisions to node $N_i$. Without loss of generality, we fix $ \mathbf{s}_i $ and shift $\mathbf{s}_j$. Denote the number of transmitting symbols $T_m$'s in $ \mathbf{s}_i $ that overlap with receiving symbols $R_m$'s in $ \mathbf{s}_j^{\tau_j} $ by $H_{i\rightarrow j}(\tau_j)$. By Lemma~\ref{lemma:pre-1}, the sum of $ H_{i\rightarrow j}(\tau_j) $ for all $ \tau_j \in \mathbb{Z}_{L} $ satisfies
\begin{equation}
\sum_{\tau_j=0}^{L-1} H_{i\rightarrow j}(\tau_j)= \alpha_i \beta_j^m.
\end{equation}
If $ \sum_{\tau_j=0}^{L-1} H_{i\rightarrow j}(\tau_j)< (k-1)L$, then there must exist at least a value of $ \tau_{j}\in \mathbb{Z}_{L}  $ such that $ H_{i\rightarrow j}(\tau_{j})<k-1 $. 
This means that with such a $\tau_{j}$, the number of $ T_{m} $'s in $ \mathbf{s}_{i} $ that overlap with $ R_{m} $'s in $ \mathbf{s}_j^{\tau_j} $ is no more than $ (k-2) $. 
However, there must exist a time offset combination of other $(k-2)$ nodes in $G_m$, such that $(k-2)$  $ T_{m} $'s in $ \mathbf{s}_{i} $ are collided.  In this case, node $ N_{i} $ would fail to transmit to node $ N_{j} $. Thus to ensure successful transmissions among all the transmitter-receiver pairs with all possible $ \bm{\tau} $, we have 
%\vspace{-0.1cm}
\begin{equation} \label{eq:bound2-e2}
\alpha_i \beta_j^m\geq (k-1)L, \text{ for any }i,j\in [K'], j \neq i, m\in [W].
\end{equation}
Summing \eqref{eq:B-LB-e1} up for $ i\in [K'] $ yields 
\begin{equation} \label{eq:bound2-e3}
K'L=\sum_{i=1}^{K'} \alpha_i+\sum_{i=1}^{K'}\sum_{r=1}^{W}\beta_i^{r}.
\end{equation}
By plugging \eqref{eq:bound2-e2} into \eqref{eq:bound2-e3}, we  obtain
\begin{equation*} 
K'L\geq \sum_{i=1}^{K'} \alpha_i+W(k -1)L\sum_{i=1}^{K'}\dfrac{1}{\alpha_i}.
\end{equation*}
Due to the fact that the harmonic mean is no more than the arithmetic mean, we  obtain
\begin{equation} \label{eq:bound2-e4}
K'L\geq \sum_{i=1}^{K'} \alpha_i+\dfrac{W(k -1)LK'^2}{\sum_{i=1}^{K'}\alpha_i}.
\end{equation}
Then \eqref{eq: B-bound2-e1} follows by the fact that the minimum value of the RHS of \eqref{eq:bound2-e4} is $ 2K'\sqrt{W(k -1)L} $.

For the case of $k=1$, we consider the lower bound on $L$ for $K'=W$ nodes. In this case, any transmitter-receiver pair should satisfy 
\begin{equation} \label{eq:bound2-e5}
\alpha_i \beta_j^m\geq L, \text{ for any }i,j\in [K'], j \neq i, m\in [W].
\end{equation}
Then by the same analysis, we have 
\begin{equation} \label{eq:bound2-e6}
K'L\geq \sum_{i=1}^{K'} \alpha_i+(K'-1)L\sum_{i=1}^{K'}\dfrac{1}{\alpha_i}.
\end{equation}
Based on \eqref{eq:bound2-e6}, we have
\begin{equation*}
L\geq 4(K'-1)=4(W-1).
\end{equation*}
This completes the proof for Theorem~\ref{thm: B-bound2}.  
\end{IEEEproof}

By combining Theorem~\ref{theorem: B-lower bound for general M} and Theorem~\ref{thm: B-bound2}, we can conclude that for an $ (M,K,L) $-schedule sequence set, when $ k=1 $,
\begin{equation} \label{eq:B-small bound}
\Omega(W,k,M,K)\geq 4(W-1),
\end{equation}
and when $ k\geq 2 $,
\begin{equation} \label{eq:B-combined bound}
\Omega(W,k,M,K)\geq  \max \left \{ \left \lceil \dfrac{8 W(k -1)^{3}  }{9k} \right\rceil, 4W(k -1) \right \}.
\end{equation}

Especially, for the even group division case with $W=M$, $k =\lfloor K/M \rfloor\geq 2$, the lower bound  is
\begin{align} 
&\Omega(W,k,M,K)\geq \notag \\
& \max \left \{ \left \lceil \dfrac{8 M(\lfloor K/M \rfloor -1)^{3}  }{9\lfloor K/M \rfloor} \right\rceil, 4M(\lfloor K/M \rfloor -1) \right \}.\label{eq:B-combined bound-even}
\end{align}

%%%%%%%%%%%%%%%%%%%%%%%%%%%%%%%%%%%%%%%%%%%%%%%%%%%%%%%

%%%%%%%%%%%%%%%%%%%%%%%%%%%%%%%%%%%%%%%%%%%%%%%%%%%%%%%

%\section{Sequence design with FDD}

%%%%%%%%%%%%%%%%%%%%%%%%%%%%%%%%%%%%%%%%%%%%%%%%%%%%%%%
\section{Construction for an $ (M,K,L) $-Schedule Sequence Set} \label{section: sequence design}

In this section, we propose a CRT-based construction for $ (M,K,L) $-schedule sequence set. For notation simplicity, this construction is called  Construction~$*$. 
Given a group division with $W$ non-empty groups $G_1,G_2, \ldots,G_{W} $ and the largest group size $ \ell $, 
%Recall that by the group-based channel allocation method, each of the $M$ groups contains $k_1 \text{ or } k_2$ nodes,  $k_1=\lceil K/M \rceil $, $k_2=\lfloor K/M  \rfloor$. In Construction~$*$, for any given $K$ and~$M$,
 we  first design a set of $W\ell$ sequences for the case that each of the $W$ groups contains exactly $\ell $ nodes, then we  obtain an $ (M,K,L) $-schedule sequence set by randomly picking $K$ sequences out of them.   

As mentioned in Section~\ref{subsec: CRT}, an array can be mapped to a sequence via the CRT correspondence~\eqref{eq: CRT mapping}, if the number of rows  and the number of columns of the array are coprime. In Construction~$*$, to design a schedule sequence $ \mathbf{s}_{i} $ for the node $ N_{i} $,   $i\in [W\ell]$, we first construct an array $ A_{i} $ consisting of $ 2W $ rows each of which is defined by a CRT-UI sequence of length $ L'=pq $. Under the construction, $ 2W $ and $ L' $ are required to be coprime with each other, so that we can map the array $ A_{i} $ to a one-dimensional sequence $ \mathbf{s}_{i} $ of length $ L=2WL' $.
Cyclically shifting $ \mathbf{s}_{i} $ by $ \tau_{i} $ is equivalent to row-wise and column-wise shifting $ A_{i} $ by corresponding time offsets. The shifted version of $ A_{i} $ is denoted by~$ A_{i}^{\tau_{i}} $.

In Construction~$*$, the $\ell$ nodes in each group are associated with a set of $\ell$ CRT-UI sequences $ \mathbf{u}_1, \mathbf{u}_2, \ldots, \mathbf{u}_{\ell} $.
Specifically, if node $ N_{i} $ is the $n$-th node in group $ G_m $, $ m\in [W] $, $ n\in [\ell] $, then each row in its array $ A_{i} $ is defined by the CRT-UI sequence $ \mathbf{u}_n $:
the positions of transmitting symbols $T_m$'s in each row are determined by ``1''s  of $ \mathbf{u}_n $.
For two nodes $N_{i}$ and $N_{j}$, if node $ N_{i} $ is the $n$-th node in  $G_{ m_1} $, and node $ N_{j} $ is the $n$-th node in  $ G_{m_2} $, $ m_1, m_2 \in [W] $, $ m_1 \neq m_2 $, $n\in [\ell]$, then all rows in their arrays $A_{i}$ and $A_{j}$ are defined by $\mathbf{u}_n$.
%If the time offsets of $N_{x_1,y}$ and $N_{x_2,y}$ are the same, i.e., $ \tau_{x_1,y}=\tau_{x_2,y} $,
If $ \tau_{i}=\tau_{j} $,
then the transmitting symbols $T_{m_1}$'s in $ A_{i}^{\tau_{i}} $ would exactly overlap with the transmitting symbols $T_{m_2}$'s in $ A_{j}^{\tau_{j}} $, which indicates that the transmissions between $N_{i}$ and $N_{j}$ would fail.
To prevent the occurrence of this case, we pre-assign rows in $A_{i}$ and $A_{j}$ with different time offsets. The effect of the pre-assigned time offsets, which is based on the auto-correlation property of CRT-UI sequences, will be explained in detail in the proof for Theorem~\ref{thm:B-feasibility}.
Here we use a simple example to illustrate the intuitive effect.
\begin{exmp} \label{exmp: pre-assigned delay offsets}
Given a binary sequence $ \mathbf{s}=[ 1\ 1\ 1\ 0\ 0] $ with length $ L=5 $, we construct two $2\times L$ arrays $ A_1 $ and $A_2$ based on $\mathbf{s}$. In each array, the first row is exactly $ \mathbf{s} $ itself, while the second row is a shifted version of $ \mathbf{s} $ with a pre-assigned time  offset. We set the time offsets as 1 and 2 for the two arrays. Then the obtained arrays  $ A_1 $ and $A_2$ are as follows,
\begin{gather*}
A_{1}=\left[
 \begin{matrix}
   \mathbf{s}\\
   \mathbf{s}^1\\
  \end{matrix}
  \right]= \left[ \setlength\arraycolsep{1.1pt}
  \begin{array}{ccccc}
1& 1& 1  & 0 &0 \\
1& 1 & 0 & 0 &1 \\
  \end{array}
\right],
A_{2}=\left[
 \begin{matrix}
   \mathbf{s}\\
   \mathbf{s}^2\\
  \end{matrix}
  \right]= \left[ \setlength\arraycolsep{1.1pt}
  \begin{array}{ccccc}
1& 1& 1  & 0 &0 \\
1 & 0 & 0 &1 &1 \\
  \end{array}
\right].
\end{gather*}
We can check that no matter how we column-wise and row-wise shift the two arrays, if the first rows in the two shifted arrays are exactly the same, then the second rows must be different, and vice versa. 
Thus for  $A_1$, even if all the ``1''s in the first or second row are collided with ``1''s in $ A_2 $, there is at least one ``1'' in the other row that can survive without collisions and can successfully match with a ``0'' in $ A_2 $.
\end{exmp}

For  $ K$ nodes and $ M $ available channels with group division parameters $W$ and $\ell$, the detailed steps of Construction~$*$  are as follows.

\begin{flushleft}
\textbf{Construction  $*$}
\end{flushleft}

\begin{enumerate}
%\item Construct a set of $ k $ CRT sequences of common period $ L'=pq $ by the CRT Construction in Section \ref{section: CRT sequences}, with Hamming weight $ w=k+1 $, $ p $ being the smallest prime that satisfies $ p\geq \text{max}\{w, 2M-2\} $, $ q $ being the smallest number coprime with $ p $ and $ 2M $, and satisfies $ q\geq 2w-1 $. These CRT sequences with generators $ 1,2,\ldots,k $ are denoted as $ s_{1}(t), s_{2}(t), \ldots, s_{k}(t)$, $t\in \mathbb{Z}_{pq}$.  The Hamming cross-correlation of each pair of them is at most 1, for all possible delay offsets.
\item Construct a set of $ \ell $  sequences $ \mathbf{u}_1, \mathbf{u}_2, \ldots, \mathbf{u}_{\ell}$ by the CRT-UI Construction, with generators $ g=1,2,\ldots,\ell $, Hamming weight $ w=\ell +1 $, $ p $ being the smallest prime that satisfies $ p\geq \text{max}\{w, 2W-2\} $, $ q $ being the smallest integer that is coprime with $ p $ and $ 2W $, and satisfies $ q\geq 2w-1 $. The common period of these CRT-UI sequences is $ L'=pq $. %The Hamming cross-correlation of any pair of them is at most 1, for all possible time offsets of these CRT-UI sequences.
\item For the node $ N_{i} $, $ i\in [W\ell] $, which is the $n$-th node in $G_m$, $ m\in [W], n\in [\ell] $, we define $ \mathbf{u}_{n}(T_{m},R_{r}) $ as the sequence obtained from $ \mathbf{u}_{n} $ by replacing ``1''s and ``0''s with $ T_{m} $'s and $ R_{r} $'s, $ r \in [W] $, respectively. Define $\mathbf{u}_{n}^{\delta_{m}} (T_{m},R_{r}) $ as the sequence obtained by cyclically shifting $ \mathbf{u}_{n}(T_{m},R_{r}) $ by a pre-assigned time offset $ \delta_{m} $, where $ \delta_{m} $ is defined as the unique integer in $ \mathbb{Z}_{L'} $ that satisfies
\begin{equation*}
\Phi_{p,q}(\delta_{m})=(m-1,0).
\end{equation*}
Stack $ \mathbf{u}_{n}(T_{m},R_{r}) $, $ \mathbf{u}_{n}^{\delta_{m}} (T_{m},R_{r}) $ together for $ r =1,2,\ldots, W$ to form a $ 2W\times L' $ array $ A_i $ as follows,
\begin{equation*}
A_{i}=\left[
 \begin{matrix}
   \mathbf{u}_{n}(T_{m},R_{1})\\
   \mathbf{u}_{n}^{\delta_{m}} (T_{m},R_{1})\\
   \mathbf{u}_{n}(T_{m},R_{2})\\
   \mathbf{u}_{n}^{\delta_{m}} (T_{m},R_{2})\\
   \vdots \\
   \mathbf{u}_{n}(T_{m},R_{W})\\
   \mathbf{u}_{n}^{\delta_{m}} (T_{m},R_{W})\\
  \end{matrix}
  \right].
\end{equation*}
\item Schedule sequence $ \mathbf{s}_{i} =[s_i(0)~ s_i(1)~ \ldots ~s_i(L-1)]$ of length $ L=2WL' $ is obtained from $ A_{i} $ via the following CRT correspondence:
$$
s_{i}(t)= A_{i}(t \text{ mod } 2W, t \text{ mod } L'),~t\in \mathbb{Z}_L.
$$
\item Randomly pick $K$ sequences out of the $ W\ell $ sequences to form an $ (M,K,L) $-schedule sequence set.
\end{enumerate}

\begin{exmp} \label{example 2}
For 4 nodes ($ K=4$)  and 2 channels ($ M=2$), we let $ W=2 $ and $ \ell=2 $, specifically, $G_1=\{N_1,N_2 \}$, $G_2=\{N_3,N_4\}$. Under this group division,  we construct 4 schedule sequences $\mathbf{s}_1, \mathbf{s}_2, \mathbf{s}_3, \mathbf{s}_4$ according to Construction~$*$.

At first, we design a set of two CRT-UI sequences $ \mathbf{u}_{1}$, $ \mathbf{u}_{2} $ with  $w=3 $, $ p=3$, $q=5$ and generators 1, 2 as follows,
\begin{align*}
&\mathbf{u}_{1}=[1\ 1\ 1\ 0\ 0\ 0\ 0\ 0\ 0\ 0\ 0\ 0\ 0\ 0\ 0]; \\
&\mathbf{u}_{2}=[1\ 0\ 0\ 0\ 0\ 0\ 0\ 1\ 0\ 0\ 0\ 1\ 0\ 0\ 0].
\end{align*}

For nodes $ N_1, N_2 \in G_1$, $ \Phi_{p,q}(\delta_{1})=(0,0) $, thus $ \delta_{1}=0 $. For nodes $ N_3, N_4 \in G_2$, $ \Phi_{p,q}(\delta_{2})=(1,0) $, thus $ \delta_{2}=10 $. For each node, we construct a $ 4\times 15 $ array as shown in \eqref{eq:exmp for construction}.  The symbols $ T_{1}, T_{2}, R_{1} $ and $ R_{2} $ are displayed in different colors in order to facilitate easier reading. Schedule sequence $ \mathbf{s}_{i} $ ($ i\in [4] $) of length $ L=60 $ is obtained from $ A_{i} $ by the mapping:
$s_{i}(t)=A_{i}(t \text{ mod }4, t\text{ mod }15)$. 
\begin{figure*}
\begin{minipage}[t]{1\linewidth}
\begin{align} \label{eq:exmp for construction}
\begin{split}
A_{1}=\begin{bmatrix}
\mathbf{u}_{1}(T_{1},R_{1})\\ \mathbf{u}_{1}^{0}(T_{1},R_{1})\\ \mathbf{u}_{1}(T_{1},R_{2})\\ \mathbf{u}_{1}^{0}(T_{1},R_{2})
\end{bmatrix}=\left[    \setlength\arraycolsep{2pt}
  \begin{array}{ccccccccccccccc}
\textcolor{r1}{T_{1}}&\textcolor{r1}{T_{1}}& \textcolor{r1}{T_{1}}&\textcolor{blue}{R_{1}}& \textcolor{blue}{R_{1}}& \textcolor{blue}{R_{1}}&\textcolor{blue}{R_{1}}&\textcolor{blue}{R_{1}}&\textcolor{blue}{R_{1}}&\textcolor{blue}{R_{1}}&\textcolor{blue}{R_{1}}&\textcolor{blue}{R_{1}}&\textcolor{blue}{R_{1}}&\textcolor{blue}{R_{1}}&\textcolor{blue}{R_{1}}\\
\textcolor{r1}{T_{1}}&\textcolor{r1}{T_{1}}& \textcolor{r1}{T_{1}}&\textcolor{blue}{R_{1}}& \textcolor{blue}{R_{1}}& \textcolor{blue}{R_{1}}&\textcolor{blue}{R_{1}}&\textcolor{blue}{R_{1}}&\textcolor{blue}{R_{1}}&\textcolor{blue}{R_{1}}&\textcolor{blue}{R_{1}}&\textcolor{blue}{R_{1}}&\textcolor{blue}{R_{1}}&\textcolor{blue}{R_{1}}&\textcolor{blue}{R_{1}}\\
\textcolor{r1}{T_{1}}&\textcolor{r1}{T_{1}}& \textcolor{r1}{T_{1}}&\textcolor{g}{R_{2}}&\textcolor{g}{R_{2}}& \textcolor{g}{R_{2}}&\textcolor{g}{R_{2}}&\textcolor{g}{R_{2}}&\textcolor{g}{R_{2}}&\textcolor{g}{R_{2}}&\textcolor{g}{R_{2}}&\textcolor{g}{R_{2}}&\textcolor{g}{R_{2}}&\textcolor{g}{R_{2}}&\textcolor{g}{R_{2}}\\
\textcolor{r1}{T_{1}}&\textcolor{r1}{T_{1}}& \textcolor{r1}{T_{1}}&\textcolor{g}{R_{2}}&\textcolor{g}{R_{2}}& \textcolor{g}{R_{2}}&\textcolor{g}{R_{2}}&\textcolor{g}{R_{2}}&\textcolor{g}{R_{2}}&\textcolor{g}{R_{2}}&\textcolor{g}{R_{2}}&\textcolor{g}{R_{2}}&\textcolor{g}{R_{2}}&\textcolor{g}{R_{2}}&\textcolor{g}{R_{2}}\\
  \end{array}
\right]   ;\\
A_{2}=\begin{bmatrix}
\mathbf{u}_{2}(T_{1},R_{1})\\ \mathbf{u}_{2}^{0}(T_{1},R_{1})\\ \mathbf{u}_{2}(T_{1},R_{2})\\ \mathbf{u}_{2}^{0}(T_{1},R_{2})
\end{bmatrix}=\left[    \setlength\arraycolsep{2pt}
  \begin{array}{ccccccccccccccc}
\textcolor{r1}{T_{1}}&\textcolor{blue}{R_{1}}& \textcolor{blue}{R_{1}}&\textcolor{blue}{R_{1}}& \textcolor{blue}{R_{1}}& \textcolor{blue}{R_{1}}&\textcolor{blue}{R_{1}}&\textcolor{r1}{T_{1}}&\textcolor{blue}{R_{1}}&\textcolor{blue}{R_{1}}&\textcolor{blue}{R_{1}}&\textcolor{r1}{T_{1}}&\textcolor{blue}{R_{1}}&\textcolor{blue}{R_{1}}&\textcolor{blue}{R_{1}}\\
\textcolor{r1}{T_{1}}&\textcolor{blue}{R_{1}}& \textcolor{blue}{R_{1}}&\textcolor{blue}{R_{1}}& \textcolor{blue}{R_{1}}& \textcolor{blue}{R_{1}}&\textcolor{blue}{R_{1}}&\textcolor{r1}{T_{1}}&\textcolor{blue}{R_{1}}&\textcolor{blue}{R_{1}}&\textcolor{blue}{R_{1}}&\textcolor{r1}{T_{1}}&\textcolor{blue}{R_{1}}&\textcolor{blue}{R_{1}}&\textcolor{blue}{R_{1}}\\
\textcolor{r1}{T_{1}}&\textcolor{g}{R_{2}}&\textcolor{g}{R_{2}}&\textcolor{g}{R_{2}}&\textcolor{g}{R_{2}}&\textcolor{g}{R_{2}}&\textcolor{g}{R_{2}}&\textcolor{r1}{T_{1}}&\textcolor{g}{R_{2}}&\textcolor{g}{R_{2}}&\textcolor{g}{R_{2}}&\textcolor{r1}{T_{1}}&\textcolor{g}{R_{2}}&\textcolor{g}{R_{2}}&\textcolor{g}{R_{2}}\\
\textcolor{r1}{T_{1}}&\textcolor{g}{R_{2}}&\textcolor{g}{R_{2}}&\textcolor{g}{R_{2}}&\textcolor{g}{R_{2}}&\textcolor{g}{R_{2}}&\textcolor{g}{R_{2}}&\textcolor{r1}{T_{1}}&\textcolor{g}{R_{2}}&\textcolor{g}{R_{2}}&\textcolor{g}{R_{2}}&\textcolor{r1}{T_{1}}&\textcolor{g}{R_{2}}&\textcolor{g}{R_{2}}&\textcolor{g}{R_{2}}\\
  \end{array}
\right] ;\\
A_{3}=\begin{bmatrix}
\mathbf{u}_{1}(T_{2},R_{1})\\ \mathbf{u}_{1}^{10}(T_{2},R_{1})\\ \mathbf{u}_{1}(T_{2},R_{2})\\ \mathbf{u}_{1}^{10}(T_{2},R_{2})
\end{bmatrix}=\left[    \setlength\arraycolsep{2pt}
  \begin{array}{ccccccccccccccc}
\textcolor{r2}{T_{2}}&\textcolor{r2}{T_{2}}& \textcolor{r2}{T_{2}}&\textcolor{blue}{R_{1}}& \textcolor{blue}{R_{1}}& \textcolor{blue}{R_{1}}&\textcolor{blue}{R_{1}}&\textcolor{blue}{R_{1}}&\textcolor{blue}{R_{1}}&\textcolor{blue}{R_{1}}&\textcolor{blue}{R_{1}}&\textcolor{blue}{R_{1}}&\textcolor{blue}{R_{1}}&\textcolor{blue}{R_{1}}&\textcolor{blue}{R_{1}}\\
\textcolor{blue}{R_{1}}&\textcolor{blue}{R_{1}}&\textcolor{blue}{R_{1}}&\textcolor{blue}{R_{1}}&\textcolor{blue}{R_{1}}& \textcolor{r2}{T_{2}}&\textcolor{r2}{T_{2}}& \textcolor{r2}{T_{2}}&\textcolor{blue}{R_{1}}& \textcolor{blue}{R_{1}}&\textcolor{blue}{R_{1}}& \textcolor{blue}{R_{1}}& \textcolor{blue}{R_{1}}&\textcolor{blue}{R_{1}}&\textcolor{blue}{R_{1}}\\
\textcolor{r2}{T_{2}}&\textcolor{r2}{T_{2}}& \textcolor{r2}{T_{2}}&\textcolor{g}{R_{2}}&\textcolor{g}{R_{2}}& \textcolor{g}{R_{2}}&\textcolor{g}{R_{2}}&\textcolor{g}{R_{2}}&\textcolor{g}{R_{2}}&\textcolor{g}{R_{2}}&\textcolor{g}{R_{2}}&\textcolor{g}{R_{2}}&\textcolor{g}{R_{2}}&\textcolor{g}{R_{2}}&\textcolor{g}{R_{2}}\\
\textcolor{g}{R_{2}}&\textcolor{g}{R_{2}}& \textcolor{g}{R_{2}}&\textcolor{g}{R_{2}}&\textcolor{g}{R_{2}}& \textcolor{r2}{T_{2}}&\textcolor{r2}{T_{2}}& \textcolor{r2}{T_{2}}&\textcolor{g}{R_{2}}&\textcolor{g}{R_{2}}&\textcolor{g}{R_{2}}&\textcolor{g}{R_{2}}&\textcolor{g}{R_{2}}&\textcolor{g}{R_{2}}&\textcolor{g}{R_{2}}
  \end{array}
\right]   ;\\
A_{4}=\begin{bmatrix}
\mathbf{u}_{2}(T_{2},R_{1})\\ \mathbf{u}_{2}^{10}(T_{2},R_{1})\\ \mathbf{u}_{2}(T_{2},R_{2})\\ \mathbf{u}_{2}^{10}(T_{2},R_{2})
\end{bmatrix}=\left[    \setlength\arraycolsep{2pt}
  \begin{array}{ccccccccccccccc}
\textcolor{r2}{T_{2}}&\textcolor{blue}{R_{1}}& \textcolor{blue}{R_{1}}&\textcolor{blue}{R_{1}}& \textcolor{blue}{R_{1}}& \textcolor{blue}{R_{1}}&\textcolor{blue}{R_{1}}&\textcolor{r2}{T_{2}}&\textcolor{blue}{R_{1}}&\textcolor{blue}{R_{1}}&\textcolor{blue}{R_{1}}&\textcolor{r2}{T_{2}}&\textcolor{blue}{R_{1}}&\textcolor{blue}{R_{1}}&\textcolor{blue}{R_{1}}\\
\textcolor{blue}{R_{1}}&\textcolor{r2}{T_{2}}& \textcolor{blue}{R_{1}}&\textcolor{blue}{R_{1}}& \textcolor{blue}{R_{1}}& \textcolor{r2}{T_{2}}&\textcolor{blue}{R_{1}}&\textcolor{blue}{R_{1}}&\textcolor{blue}{R_{1}}&\textcolor{blue}{R_{1}}&\textcolor{blue}{R_{1}}&\textcolor{blue}{R_{1}}&\textcolor{r2}{T_{2}}&\textcolor{blue}{R_{1}}&\textcolor{blue}{R_{1}}\\
\textcolor{r2}{T_{2}}&\textcolor{g}{R_{2}}&\textcolor{g}{R_{2}}&\textcolor{g}{R_{2}}&\textcolor{g}{R_{2}}&\textcolor{g}{R_{2}}&\textcolor{g}{R_{2}}&\textcolor{r2}{T_{2}}&\textcolor{g}{R_{2}}&\textcolor{g}{R_{2}}&\textcolor{g}{R_{2}}&\textcolor{r2}{T_{2}}&\textcolor{g}{R_{2}}&\textcolor{g}{R_{2}}&\textcolor{g}{R_{2}}\\
\textcolor{g}{R_{2}}&\textcolor{r2}{T_{2}}& \textcolor{g}{R_{2}}&\textcolor{g}{R_{2}}& \textcolor{g}{R_{2}}&\textcolor{r2}{T_{2}}&\textcolor{g}{R_{2}}&\textcolor{g}{R_{2}}&\textcolor{g}{R_{2}}&\textcolor{g}{R_{2}}&\textcolor{g}{R_{2}}&\textcolor{g}{R_{2}}&\textcolor{r2}{T_{2}}&\textcolor{g}{R_{2}}&\textcolor{g}{R_{2}}
  \end{array}
\right]   .
\end{split}
\end{align}
\end{minipage}
\end{figure*}

%We can check that the sequence set $\{ \mathbf{s}_{i} :  i\in \mathbb{I}_4 \}$ is a $ (2,4,60)_B $-schedule sequence set.
\end{exmp}

\begin{theorem} \label{thm:B-feasibility}
A sequence set $\{ \mathbf{s}_{i} :  i\in [K]\}$ obtained by Construction~$*$  is an $ (M,K,L) $-schedule sequence set.
\end{theorem}

%Please see Appendix~\ref{appendix B-2} for the proof of Theorem~\ref{thm:B-feasibility}.
\begin{IEEEproof}
We consider the transmission from $ N_{i} $ to  $ N_{j} $, for $i,j \in [K]$, $i\neq j$. Assume that $N_i$ is the $n$-th node in group $G_m$, and $N_j$ is the $y$-th node in group $G_x$, for $m,x\in [W]$, $n,y\in [\ell]$.
Since the transmission is successful only when  $N_{j}$ receives on channel~$m$ and at the same time  $ N_{i} $ transmits on channel~$m$ without colliding with other nodes, thus we need to consider the positions of $ R_m $'s in $ \mathbf{s}_{j}^{\tau_{j}} $, or equivalently, $ A_{j}^{\tau_{j}} $.
By Construction~$*$, there are two rows containing $R_m$'s in $A_{j}$: row $2(m-1)$ which is $ \mathbf{u}_y(T_x,R_m) $ and row $(2m-1)$ which is $ \mathbf{u}_y^{\delta_x}(T_x,R_m) $. 
After row-wise and column-wise shifting, there are still only two rows containing $R_m$'s in $A_{j}^{\tau_{j}}$.
The row indices of these two rows are denoted by~$ \eta_{1} $ and~$ \eta_{2} $, where $ \eta_{1}, \eta_{2}\in \mathbb{Z}_{2W} $. Row~$ \eta_{1} $ and row~$ \eta_{2} $ in $ A_{i}^{\tau_i} $ and $ A_{j}^{\tau_{j}} $ are shown as follows,
%The symbol $ R_{*} $ in $ A_{i,j}^{\tau_{i,j}} $ just means a receiving symbol and we don't need to care about its value. The delay offsets $ \theta_{1},  \theta_{2}$ in $ A_{i,j}^{\tau_{i,j}} $  depend on $ \tau_{i,j} $ and $ \delta_{i} $. While the delay offsets $ \phi_{1}, \phi_{2} $ in $ A_{m,n}^{\tau_{m,n}} $ depend on $ \tau_{m,n} $ and $ \delta_{m} $.
\begin{align*}
A_{i}^{\tau_{i}}=\begin{array}{|c|}
\hline
\vdots\\
\hline
\text{row } \eta_{1}: \mathbf{u}_{n}^{\theta_{1}}(T_{m},R_{*})  \\
\hline
\text{row } \eta_{2}:  \mathbf{u}_{n}^{\theta_{2}}(T_{m},R_{*}) \\
\hline
\vdots \\
\hline
\end{array}, 
A_{j}^{\tau_{j}}=\begin{array}{|c|}
\hline
\vdots\\
\hline
\text{row } \eta_{1}: \mathbf{u}_{y}^{\phi_{1}}(T_{x},R_{m}) \\
\hline
\text{row } \eta_{2}: \mathbf{u}_{y}^{\phi_{2}}(T_{x},R_{m})  \\
\hline
\vdots \\
\hline
\end{array}.
\end{align*}

Note that in row $ \eta_{1} $ and row $ \eta_{2} $ of $ A_{j}^{\tau_{j}} $, $ R_{m} $ positions are defined by the CRT-UI sequence $ \mathbf{u}_{y} $ with time offsets $ \phi_{1} $ and $ \phi_{2} $ respectively. The values of $ \phi_{1}, \phi_{2} $ depend on $ \tau_{j} $, and satisfy $ \vert \phi_{1}-\phi_{2} \vert=\delta_{x} $. In row $ \eta_{1} $ and row $ \eta_{2} $ of $ A_{i}^{\tau_{i}} $, $ T_{m} $ positions are defined by the CRT-UI sequence $ \mathbf{u}_{n} $ with time offsets $ \theta_{1} $ and $ \theta_{2} $ respectively. The values of $ \theta_{1}, \theta_{2}$ depend on $ \tau_{i} $, and satisfy $ \vert \theta_{1}-\theta_{2} \vert=\delta_{m} $. The symbol $ R_{*} $ in $ A_{i}^{\tau_{i}} $ indicates a receiving symbol and the exact channel number is immaterial for the discussion.

We denote the number of collision-free $T_m$'s in row $\eta_1$ (resp. $\eta_2$) of $A_{i}^{\tau_{i}}$ that overlap with an $R_m$, instead of a $T_x$, in row $\eta_1$ (resp. $\eta_2$) of $A_{j}^{\tau_{j}}$ by $\mathcal{N}_{\eta_1}$ (resp. $\mathcal{N}_{\eta_2}$);
%We denote the number of $T_i$'s in row $\eta_1$ ($\eta_2$) of $A_{i,j}^{\tau_{i,j}}$ that overlap with a $R_i$ in row $\eta_1$ ($\eta_2$) of $A_{m,n}^{\tau_{m,n}}$ by $\mathcal{N}_{\eta_1}$ ($\mathcal{N}_{\eta_2}$).
denote the number of $T_m$'s in row $\eta_1$ (resp. $\eta_2$) of $A_{i}^{\tau_{i}}$ that overlap with a $T_x$ in row $\eta_1$ (resp. $\eta_2$) of $A_{j}^{\tau_{j}}$ by $\mathcal{N}_{\eta_1}^{o}$ (resp. $\mathcal{N}_{\eta_2}^{o}$);
and denote the number of $ T_m $'s in row $\eta_1$ (resp. $\eta_2$) of $A_{i}^{\tau_{i}}$ that collide with $T_m$'s in other arrays of $G_m$ by $\mathcal{N}_{\eta_1}^{c}$ (resp. $\mathcal{N}_{\eta_2}^{c}$). It is obvious that 
\begin{equation} \label{eq:number of successful times}
\mathcal{N}_{\eta_1}\geq w-\mathcal{N}_{\eta_1}^{o} - \mathcal{N}_{\eta_1}^{c},~ \mathcal{N}_{\eta_2}\geq w-\mathcal{N}_{\eta_2}^{o} - \mathcal{N}_{\eta_2}^{c},
\end{equation}
where $w=\ell+1$.

Next we verify whether the following condition can be satisfied: for all possible $ \bm{\tau} $, in row $ \eta_{1} $ or row $ \eta_{2} $ of $ A_{i}^{\tau_{i}} $, there is at least one collision-free $ T_{m} $ that overlaps with an $ R_{m}  $ in $ A_{j}^{\tau_{j}} $. That is, $ \max(\mathcal{N}_{\eta_1},\mathcal{N}_{\eta_2})\geq 1 $ for all possible~$ \bm{\tau} $. To verify this, we  need to check the following three~cases.
In each case, we have $\mathcal{N}_{\eta_1}^{c}, \mathcal{N}_{\eta_2}^c\leq \ell -1$ for all possible $ \bm{\tau} $. This is because by Construction~$*$, rows of the arrays in $G_m$ are determined by $|G_m|$ CRT-UI sequences from $ \mathbf{u}_1, \mathbf{u}_2, \ldots, \mathbf{u}_{\ell} $, which have the property that the Hamming cross-correlation of any pair of them is at most~1, no matter how we cyclically shift them. %Thus in either row~$\eta_1$ or row~$\eta_2$ of $ A_{i,j}^{\tau_{i,j}} $,  there are at most $(k-1)$ $T_i$'s that would collide with $T_i$'s in other $(k-1)$ arrays of group~$i$.

%When $ j\neq n $, $ \mathcal{N}_{\eta_1}^{o}, \mathcal{N}_{\eta_2}^{o} \leq 1$. This is because the values of $ \mathcal{N}_{\eta_1}^{o}, \mathcal{N}_{\eta_2}^{o} $ are determined by the Hamming cross-correlation of $ s_j(t) $ and $s_n(t)$, which is no more than~1 for all possible $\theta_1, \theta_2$ and $\phi_1, \phi_2$.  While when $ j=n $, $i\neq m$, their values are determined by the auto-correlation of $ s_j(t) $.

\begin{enumerate}%{Case 1}
\item  Both the transmitter $ N_{i} $ and the receiver $ N_{j} $ come from the same group, i.e., $ m=x$, $ n\neq y $.
In this case, $ \mathcal{N}_{\eta_1}^{o} +\mathcal{N}_{\eta_1}^{c} \leq \ell -1 $, $ \mathcal{N}_{\eta_2}^{o} +\mathcal{N}_{\eta_2}^{c} \leq \ell-1 $. Then by \eqref{eq:number of successful times}, we have $ \mathcal{N}_{\eta_1}, \mathcal{N}_{\eta_2}\geq 2 $.

\item The transmitter $ N_{i} $ and the receiver $ N_{j} $ come from different groups and $n\neq y $.
In this case, $ \mathcal{N}_{\eta_1}^{o}, \mathcal{N}_{\eta_2}^{o} \leq 1$. This is because the values of $ \mathcal{N}_{\eta_1}^{o}, \mathcal{N}_{\eta_2}^{o} $ are determined by the Hamming cross-correlation of $ \mathbf{u}_n $ and $\mathbf{u}_y$, which is no more than~1 for all possible $\theta_1, \theta_2$ and $\phi_1, \phi_2$.
 Then by \eqref{eq:number of successful times} and the fact that $ \mathcal{N}_{\eta_1}^{c}, \mathcal{N}_{\eta_2}^{c} \leq \ell-1 $, we have $ \mathcal{N}_{\eta_1}, \mathcal{N}_{\eta_2}\geq 1 $.

%In this case, among the $w=k+1$ $T_i$'s in row  $\eta_1$ ($ \eta_2 $) of $A_{i,j}^{\tau_{i,j}}$, there are at most $(k-1)$ $T_i$'s that would collide with $T_i$'s in other $(k-1)$ arrays of group~$i$, and at most one $T_i$ that would overlap with a $T_m$ in $A_{m,n}^{\tau_{m,n}}$. Therefore, in either row $ \eta_{1} $ or row $ \eta_{2} $ in $ A_{i,j}^{\tau_{i,j}} $, there is at least one collision-free $ T_{i} $ that overlaps with an $ R_{i} $ in $ A_{m,n}^{\tau_{m,n}} $.

\item %The transmitter $ N_{i,j} $ and the receiver $ N_{m,n} $ come from different groups but correspond to an identical CRT-UI sequence, i.e., $ i\neq m$,  $j= n $.
The transmitter $ N_{i} $ and the receiver $ N_{j} $ come from different groups and $n= y $.
%In this case, each row of $ A_{i,j}^{\tau_{i,j}} $ and $ A_{m,n}^{\tau_{m,n}} $ is obtained from the same CRT-UI sequence $ s_{j}(t) $.  Then the number of $ T_{i} $'s in row $ \eta_{1} $ ($\eta_{2}$) of $ A_{i,j}^{\tau_{i,j}} $ that overlap with an $ R_{i} $ in row $ \eta_{1} $ ($\eta_{2}$) of $ A_{m,n}^{\tau_{m,n}} $ is determined by the auto-correlation value of $ s_{j}(t) $.
%In this case, $ \mathcal{N}_{\eta_1}^{c}, \mathcal{N}_{\eta_2}^{c} \leq k-1 $,  and $ \min(\mathcal{N}_{\eta_1}^{o}, \mathcal{N}_{\eta_2}^{o}) \leq 1$ according to the following Proposition \ref{proposition:time offset}. Then by \eqref{eq:number of successful times}, we have $ \max(\mathcal{N}_{\eta_1},\mathcal{N}_{\eta_2})\geq 1 $.
In this case, the values of $ \mathcal{N}_{\eta_1}^{o}, \mathcal{N}_{\eta_2}^{o} $ are determined by the auto-correlation of $ \mathbf{u}_n $. By the following Proposition \ref{proposition: B-delay offset}, we have $ \min(\mathcal{N}_{\eta_1}^{o}, \mathcal{N}_{\eta_2}^{o}) \leq 1$ for all possible $\theta_1, \theta_2$ and $\phi_1, \phi_2$. Then by \eqref{eq:number of successful times} and the fact that $ \mathcal{N}_{\eta_1}^{c}, \mathcal{N}_{\eta_2}^{c} \leq \ell-1 $, we have $ \max(\mathcal{N}_{\eta_1},\mathcal{N}_{\eta_2})\geq 1 $.
\end{enumerate}

In summary,  $ \max(\mathcal{N}_{\eta_1},\mathcal{N}_{\eta_2})\geq 1 $ for all possible~$\bm{\tau}$. That is, the sequence set $\{ \mathbf{s}_{i} :  i\in [K] \}$ obtained by Construction~$*$  can guarantee at least one collision-free transmission from $N_{i}$ to $N_{j}$, for any $ i, j \in [K] $, $ i\neq j$, and for all possible~$\bm{\tau}$. Therefore, it is an $ (M,K,L) $-schedule sequence~set.
\end{IEEEproof}

\begin{proposition} \label{proposition: B-delay offset}
If $ m\neq x$, $ n= y $, then  $ \min(\mathcal{N}_{\eta_1}^{o}, \mathcal{N}_{\eta_2}^{o}) \leq 1$ for all possible $\theta_1, \theta_2$ and $\phi_1, \phi_2$.
\end{proposition}

\begin{IEEEproof}
By Construction~$*$, 
$ \vert \theta_{1}-\theta_{2} \vert=\delta_{m}, \vert \phi_{1}-\phi_{2} \vert=\delta_{x}.$
 Denote the time offset between row $ \eta_{1} $ in $ A_{i}^{\tau_{i}} $ and row $ \eta_{1} $ in $ A_{j}^{\tau_j} $ by $ \tau_{i,j,1} $, $\tau_{i,j,1}=\theta_{1} - \phi_{1}$. Denote the time offset between row $ \eta_{2} $ in $ A_{i}^{\tau_{i}} $ and row $ \eta_{2} $ in $ A_{j}^{\tau_j} $ by $ \tau_{i,j,2} $, $\tau_{i,j,2}=\theta_{2} - \phi_{2}$. It follows that
\begin{equation} \label{eq:relationship between delay offsets}
\tau_{i,j,2}=\tau_{i,j,1}\pm(\delta_m\pm\delta_x).
\end{equation}

Next we first analyze the value of $\mathcal{N}_{\eta_1}^{o}$, which equals the Hamming auto-correlation between $\mathbf{u}_n$ and $\mathbf{u}_n^{\tau_{i,j,1}}$, $H_{n,n}(\tau_{i,j,1})$, in the following two cases. %That is, $ \mathcal{N}_{\eta_1}^{o}= H_{j,j}(\tau_{i,m,1})$.
\begin{enumerate}
\item $\Phi_{p,q}(\tau_{i,j,1})\neq (n,1)a $, for $ a \in \{0,\pm 1,\ldots,\pm (w-2)  \} $. In this case, $\mathcal{N}_{\eta_1}^{o}=0 \text{ or } 1$ since we have $ H_{n,n}(\tau_{i,j,1})=0 \text{ or } 1$ by Lemma~\ref{auto-correlation}. 
\item $\Phi_{p,q}(\tau_{i,j,1})= (n,1)a $, for $ a \in \{0,\pm 1,\ldots,\pm (w-2)  \} $. In this case, $\mathcal{N}_{\eta_1}^{o}=H_{n,n}(\tau_{i,j,1})\geq 2$. Then we analyze the value of $ \mathcal{N}_{\eta_2}^{o} $, which equals the Hamming auto-correlation between $\mathbf{u}_n$ and $\mathbf{u}_n^{\tau_{i,j,2}}$, $H_{n,n}(\tau_{i,j,2})$. We prove $ \mathcal{N}_{\eta_2}^{o}\leq 1 $ by contradiction as follows.
Assume that $ H_{n,n}(\tau_{i,j,2})\geq 2 $. It implies that $  \Phi_{p,q}(\tau_{i,j,2})= (n,1)b $, for $ b \in \{0,\pm 1, \ldots,\pm (w-2)\} $.   Let $ c=a-b $, $ c\in \{0,\pm 1, \ldots,\pm 2(w-2)\} $. Then we discuss whether this assumption can hold in the following cases indicated by~\eqref{eq:relationship between delay offsets}:
\begin{enumerate}
\item $ \tau_{i,j,2}=\tau_{i,j,1}-(\delta_m-\delta_x) $. In this case,
\begin{align*}
 \Phi_{p,q}(\delta_m-\delta_x)&=\Phi_{p,q}(\tau_{i,j,1}-\tau_{i,j,2})\\
 &=(nc \text{ mod } p, c \text{ mod } q).
\end{align*}
By Construction~$*$, $ \Phi_{p,q}(\delta_{m})=(m-1,0) $,  $ \Phi_{p,q}(\delta_{x})=(x-1,0) $, then we have $ \Phi_{p,q}(\delta_m-\delta_x)=((m-x) \text{ mod } p,0)$. Thus \begin{align}
m-x  & \equiv  nc \text{ mod } p, \label{eq:offset e1} \\
0  & \equiv c \text{ mod } q.  \label{eq:offset e2}
\end{align}
Since $ q\geq 2w-1> |c| $, then \eqref{eq:offset e2} implies $ c=0 $.
However, \eqref{eq:offset e1} cannot hold with $ c=0 $ because $ m-x \in \{\pm 1, \pm 2,\ldots, \pm (W-1) \} $ and $ p\geq 2W-2$.
\item $ \tau_{i,j,2}=\tau_{i,j,1}-(\delta_m+\delta_x) $. In this case,
\begin{align*}
\Phi_{p,q}(\delta_m+\delta_x)=(nc\text{ mod } p, c\text{ mod } q).
\end{align*}
By Construction~$*$, $ \Phi_{p,q}(\delta_m+\delta_x)=((m+x-2)\text{ mod } p,0)$, thus
\begin{align}
m+x-2  & \equiv  nc \text{ mod } p, \label{eq:offset e4} \\
0  & \equiv c \text{ mod } q.  \label{eq:offset e5}
\end{align}
Again, \eqref{eq:offset e5} implies $ c=0 $. However, \eqref{eq:offset e4} cannot hold with $ c=0 $ since $ m+x-2 \in \{1,2,\ldots, 2W-3 \} $ and $ p\geq 2W-2$.
\item $ \tau_{i,j,2}=\tau_{i,j,1}+(\delta_m\pm\delta_x) $. By the same analysis for case (a) and  (b), we can also derive that the assumption that $  H_{n,n}(\tau_{i,j,2})\geq 2 $ cannot hold.
\end{enumerate}
Therefore, we  conclude that if $\mathcal{N}_{\eta_1}^{o}\geq 2$, then  $ \mathcal{N}_{\eta_2}^{o}\leq 1 $.
\end{enumerate}
This completes the proof for Proposition~\ref{proposition: B-delay offset}.
\end{IEEEproof}

\section{Discussion on Period $ L $ under Even Group Division} \label{section: comparison}
In this section, we discuss the sequence period obtained by Construction~$*$ under even group division.
%, in which the largest group size~$\ell$ is minimized for fixed $W$. %The reason why we consider even group division is because that  in this case. 
Given~$K$ and $M$, we consider the optimal value of $W$ that can minimize period $L$. We propose the following algorithm: Define
%Note that by  Construction~$*$ under even group division, one may not obtain sequence sets with the shortest period if $W=M$. Instead, we propose the following algorithm: Define
\begin{equation} \label{eq:B-TDD-threshold}
M'=\left\lceil \sqrt{\dfrac{K}{2}+\dfrac{9}{16}}+\dfrac{3}{4}\right\rceil;
\end{equation}
if $ M> M' $, then let $W=M'$, otherwise let $W=M$. 

We provide an intuitive argument for this algorithm as follows. 
%Given $M$, we first assume $W=M$. 
By Construction~$*$,  $ w=\ell +1 =\left \lceil K/W\right \rceil +1$, $ p $ is the smallest prime that satisfies $ p\geq \text{max}\{w,2W-2\} $. To simplify the following discussion, we assume $\lceil K/W \rceil \approx K/W$. 
\begin{enumerate}
\item $ M \leq M' $. In this case, $W \leq M'$, then $ w\geq 2W-2 $, thus $ p\geq w $. When $ w $ is a prime, and $ (2w-1) $ is coprime with $ 2W $ and $ w $, we set $ p=w $ and $ q=2w-1 $, then
\begin{align} 
L=2Wpq=\dfrac{4K^2}{W}+6K+2W. \label{eq:sequence length 1}
\end{align}
By \eqref{eq:sequence length 1},  $ L $ is a decreasing function of $ W $ when $K$ is fixed and $ W\leq M' $. Therefore, in the case of $M\leq M'$, $L$ is minimized when $W=M$. 
\item $ M >M' $. We consider $W\geq M'$. In this case, $ p\geq 2W-2 $. When $ (2W-1) $ is a prime, and $ (2w-1) $ is coprime with $ 2W $ and $(2W-1)$, we set $ p=2W-1 $ and $ q=2w-1$, then
\begin{align} 
L=2Wpq=4W^2+2W(4K-1)-4K.\label{eq:sequence length 2}
\end{align}
We can see that $ L $ in \eqref{eq:sequence length 2} is an increasing function of $ W $ when $K$ is fixed. Therefore, in the case of $M>M'$, $L$ is minimized when $W=M'$.
\end{enumerate}

%Therefore, when $ M> M' $, it is better to use only $ M' $ of $ M $ channels, that is, let $W=M'$.  %Moreover, by comparing \eqref{eq:sequence length 1} with the shortest known period for single channel (see \eqref{eq:single channel}), we can observe that when $M=2$, it is better to use only one channel. 

For general $K$ and $ M\leq M' $, under even division with $W=M$, we have the following result for $L$ obtained by Construction~$*$.

%\begin{comment}
\begin{proposition} \label{prop: B-length}
Under even group division with $W=M \leq M' $,  there exists a schedule sequence set by Construction~$*$  with sequence period 
\begin{equation} \label{eq: B-length}
L\leq 2M\left( 2\left \lceil \dfrac{K}{M} \right \rceil+2\right)\left( 4 \left \lceil \dfrac{K}{M} \right \rceil+2\right).
\end{equation}
\end{proposition}
\begin{IEEEproof}
By Construction~$*$,  $w=\ell+1=\left \lceil K/M \right \rceil +1 \geq 2M-2$, $p$ is the smallest prime that satisfies $p\geq w$. By Bertrand's postulate, we have $ p< 2w $. We set $q$ as the smallest prime that satisfies $q\geq 2w-1$. It is obvious that such a $q$ is coprime with $2M$ and $p$. Again by Bertrand's postulate,  we have $q\leq 4w-2$.
Thus, the obtained period $ L=2Mpq $ satisfies~\eqref{eq: B-length}.
% In this setting, we can obtain sequence period that satisfy~\eqref{eq: B-length}. 
%Both of the two integers $(2w-1)$ and $(2w+1)$ are coprime with $p$ since $p$ is a prime and $ w\leq p<2w $.  We prove that $(2w-1)$ and $(2w+1)$ have no common non-1 factors by contradiction: Assume $2w-1=xa$, $2w+1=ya$, $a> 2$. However, $2w+1-(2w-1)=(y-x)a=2$. Thus $(2w-1)$ and $(2w+1)$ are coprime with each other. Therefore, if $(2w-1)$ is coprime with $2M$, then we can set $q=2w-1$; Otherwise we can set $q=2w+1$.
\end{IEEEproof}
%\end{comment}
%\vspace{-3mm}

%\vspace{-7mm}

By comparing~\eqref{eq: B-length} with the lower bound~\eqref{eq:B-combined bound-even} obtained in Section~\ref{sec:B-lower bound}, we can observe that under even group division with $W=M\leq M'$, the period under Construction~$*$  can achieve the same order in~$ K $ and~$ M $ as the lower bound, that is, $O(K^2/M)$.
To illustrate the gap between the period under Construction~$*$ and the lower bound, we list their ratios for some $K$ and $M$ in Table~\ref{table: BTAT/bound}. For example, when $K=70$ and $M=4$, the shortest period by Construction~$*$ is $L=5624$, and the lower bound by \eqref{eq:B-combined bound-even} is $L\geq 857$, then the ratio between them is $ 5624/857 \approx 6.56 $. We will try to reduce the gap in future work.

Moreover, we can see from \eqref{eq: B-length} and the achievable length~\eqref{eq:single channel-e1} for the single channel case that under even group division with $ W=M\leq M' $, the period under Construction~$*$  can achieve an asymptotic reduction by a factor of~$ M $. 

\begin{table}[htbp]
\caption{Ratio of the shortest $L$ under Construction~$*$ divided by the lower bound, under even group division with $W=M\leq M'$.}
\label{table: BTAT/bound}
\center
\scalebox{0.8}{
\begin{tabular}{|c|c|c|c|c|c|c|c|c|c|c|c|}
\hline
\diagbox{$M$}{$ K $} & 60 & 70 & 80 & 90 & 100 & 110 & 120  &130 &140 &150\\
\hline
2& 5.23 & 5.26 &5.04 &	5.08 &	5.12&	5.15 &	4.85 &	4.9 &	4.8 &	4.97
\\
\hline
3&  6.18 &	6.9 &	5.97 &	5.23 &	5.95 &	5.96 &	5.16 &	5.46 &	5.72 &	5.12 \\
\hline
4&  6.48 &	6.56 &	6.18 &	7.28 &	6.02 &	5.71 &5.23 &	5.99 &	5.26 &	5.63 \\
\hline
5&7.12 &	7.06 &	5.98 &	5.79 &	6.18 & 5.78 &	6.3 &5.75 &	5.29 &	5.23 \\
\hline
\end{tabular}}
\end{table}
\vspace{-0.5cm}

\section{Random Schemes} \label{sec:random}
In this section, we analyze the optimal transmitting and receiving probabilities for two random  schemes, given $K$ nodes and $W$ of $M$ channels being employed, $1\leq W\leq M$.

\subsection{General random scheme}
In the general random scheme, there is no concept of groups. For any node, at a time slot, it transmits on any one of the~$W$ channels with probability $ p_{a} $ and receives on any one of the~$W$ channels with probability~$ q_{a} $. 
The values of $ p_{a} $ and $ q_a $ satisfy $ 0<p_a <1 $, $0<q_a<1$, and $W(p_{a}+q_{a})=1$. 
The probability for any node to successfully receive a packet from another node in a time slot is
\begin{equation} \label{eq:random-e4}
P_{a}=Wp_{a}q_{a}(1-p_{a})^{K-2} =p_{a}(1-Wp_{a})(1-p_{a})^{K-2}.
\end{equation}

\subsection{Assignment T based random scheme}
In the Assignment T based random scheme, we divide $ K $ nodes into $ W $ non-empty groups. The $ |G_m| $ nodes belong to group~$ G_m $ can only transmit on channel~$ m $, for $ m\in [W] $. For node $ N_{i} \in G_m $, $ i\in [K] $, $m\in [W]$, at a time slot, it transmits on channel $ m $ with probability $ p_{b} $,  receives on channel $ m $ with probability $ q_{1} $, and receives on any other channel with probability $ q_{2} $. 
The values of $ p_b, q_1, q_2 $ satisfy $0<p_b<1$, $0<q_1<1$, $0<q_2<1$, and 
\begin{equation} \label{eq:AT-rand-e1}
p_{b}+q_{1}+(W-1)q_{2}=1.
\end{equation}
The probability for node $ N_{i} $ to successfully transmit a packet to a node in  $G_m$ in a time slot is
\begin{equation} \label{eq:AT-rand-e2}
P_{\alpha}=p_{b}q_{1}(1-p_{b})^{|G_m|-2}.
\end{equation}
The probability for node $ N_{i} $ to successfully transmit a packet to a node in another group in a time slot is
\begin{equation} \label{eq:AT-rand-e3}
P_{\beta}=p_{b}q_{2}(1-p_{b})^{|G_m|-1}.
\end{equation}

%\subsubsection{ $L_{rand}$ under same power consumption as the sequence scheme}
%We compare $L$ and $L_{rand}$ when sequence scheme and Assignment T random scheme are under the same power assumtion. 

\subsection{Optimized random scheme} \label{sec: B-optimal random}
In this section, we try to optimize the two random schemes.
% and compare the corresponding $L_{rand}$ with $L$.
At first, for the general random scheme, we find from \eqref{eq:random-e4} that for any given $K$ and $p_a$, $P_a$ monotonically decreases as $W$ increases. 
%Thus for the general random scheme, using a single channel is optimal for maximizing this probability. 
Next we consider $P_{\alpha}$ (see \eqref{eq:AT-rand-e2}) and $P_{\beta}$ (see \eqref{eq:AT-rand-e3}) in the Assignment T based random scheme. To simplify the discussion, we assume  $P_{\alpha}=P_{\beta}$ and $|G_m|=K/W$ for any $m\in [W]$. Then by \eqref{eq:AT-rand-e1}, \eqref{eq:AT-rand-e2} and \eqref{eq:AT-rand-e3}, we have
\begin{equation} \label{eq:AT-rand-e5}
  P_{\beta}=p_b(1-p_b)^{K/W} /(W-p_b).
  \end{equation}
  We can  observe from \eqref{eq:AT-rand-e5} that  for any given~$K$ and $p_b$, $ P_{\beta} $ also monotonically decreases as $W$ increases.
Thus for both of the general random scheme and the Assignment T based random scheme, using only one channel, i.e., $W=1$, is optimal for maximizing $ P_a $ or $P_{\beta}$. This implies that the two random schemes cannot efficiently make use of the multi-channel resources. 
%the probability for an arbitrary pair to transmit successfully at a time slot. 

With only one channel, the two random schemes are equivalent. For each node at each time slot, it transmits on this channel with probability~$p$ and receives on this channel with probability~$(1-p)$. 
%Then the probability for an arbitrary pair to transmit successfully at a time slot,
Then the probability for a node to receive from one of its $(K-1)$ neighboring nodes successfully at a time slot, denoted by $P$, equals $p(1-p)^{K-1}$. By taking derivative of~$P$ with respect to~$p$, we obtain that~$ P $ attains its maximum value when $ p =1/K$. Thus the optimal transmitting probability for the two random schemes with one channel~is $$p^*=1/K. $$ The corresponding $P$ is denoted by $P^*$,
\begin{equation} \label{eq: B-optimal-random}
P^*=\dfrac{(K-1)^{K-1}}{K^K}. 
\end{equation}

\section{Comparison between Sequence Scheme and Random Scheme}\label{sec:B-compare}
In this section, we compare the frame length and broadcast completion time under our proposed sequence scheme with those under the optimized random scheme. The schedule sequences employed are obtained by Construction $*$ under even group division. 

\subsection{Frame length} 
%In this section, we analyze the optimal transmitting and receiving probabilities for two random  schemes, in order to minimize the frame length.
%A key parameter that affects the performance of the scheduling schemes is the frame length. 
%The major performance metric is the frame length. 
First, we explain how frame length is defined for random schemes.
While it is possible to find sequence
schemes that can ensure each node has at least one successful broadcast per frame, it is impossible to provide such guarantee
for the random schemes.  Using a high value for frame length would strengthen the guarantee but weaken the performance of
the random schemes.   In order to determine a fair frame length value for the random schemes, we adopt the following argument.
We assume that all the nodes start at time $ t=0 $ without any offset to render the analysis manageable.  Let $X$ be the time required for each of the $K$ nodes to broadcast a packet to all other nodes  at least once, which is also the time required for each node to receive a packet from each other node at least once.  Motivated by the least required reliability for URLLC \cite{bennis2018ultrareliable}, we set the probability $P(X \leq L_{rand}) $ as $99.999\%$.  We then set the frame length to be $L_{rand}$.  
% Note that with non-zero offsets among the nodes, the probability just accounts for the event that all nodes have a successful transmission without regard to the content of the successful~packet.  

Let $ X_{i} $ be the time required for node $ N_{i} $ to receive a packet from each other node at least once, for $i\in [K]$. By definition, we have $$X=\max_{i\in [K] } X_{i}.$$
To simplify calculation, we follow the assumption in \cite{wu2014safety}, that is, $X_{i}$'s are assumed to be independent for all $ i\in [K] $. Then for any $\ell \geq 0$, $ P(X\leq \ell) $ can be obtained from $ P(X_{i}\leq \ell) $~by
\begin{equation} \label{eq: I-group-delay}
P(X\leq \ell)=\prod_{i=1,...,K}P(X_{i}\leq \ell).
\end{equation} 
 We  analyze $ P(X_{i}\leq \ell) $ by using results in the coupon collector's problem \cite{adler2003coupon}. 
We use $\mathcal{E}_i^j$ to denote the event that at a time slot, the $j$-th  of the $ (K-1) $ neighboring nodes of $N_{i}$ successfully transmits a packet to node $N_{i}$, for $j=1,2,\ldots,K-1$. We  use $\mathcal{E}_i^0$ to denote the event that at a time slot, none of the $ (K-1) $ events $ \mathcal{E}_i^1 , \mathcal{E}_i^2, \ldots, \mathcal{E}_i^{K-1} $ happens. Then by definition, $X_{i}$ is exactly the time slots needed for the $ (K-1) $ events  $ \mathcal{E}_i^1 , \mathcal{E}_i^2, \ldots, \mathcal{E}_i^{K-1} $  to happen at least once. 
Now consider a coupon collector's problem: in a container indexed by $i$,  there are $K$ coupons which are randomly drawn one by one with replacement. 
Among these $K$ coupons, there are $ (K-1) $ different coupons corresponding to events $ \mathcal{E}_i^1 , \mathcal{E}_i^2, \ldots, \mathcal{E}_i^{K-1} $, and a null coupon corresponding to the event $ \mathcal{E}_i^0 $. Let~$Y_i$ be the time required to get a collection of $ (K-1) $ different coupons. Then $X_{i}$ has the same distribution as $Y_i$. That is, for any $\ell\geq 0$, 
\begin{equation} \label{eq: I-group-delay-e2}
P(X_{i}\leq \ell) = P(Y_i \leq \ell). 
\end{equation}

% can be regarded as the time required to get a collection of $ (K-1) $ different coupons.
Let $\bm{p}_i=[p(\mathcal{E}_i^0)~p(\mathcal{E}_i^1)~ \ldots ~p(\mathcal{E}_i^{K-1}) ],$
 where $ p(\mathcal{E}_i^j) $ denotes the probability that the event $\mathcal{E}_i^j$ happens at a time slot, which  also denotes the probability that coupon $j$ is drawn at a time slot, for $j=0,1,\ldots,K-1$. Since the distribution of $Y_i$ depends on $\bm{p}_i$, we will use the notation~$Y_i(\bm{p}_i)$. Let $\bm{P}=[\bm{p}_1 \ \bm{p}_2 \ \ldots \ \bm{p}_K]$. Since the value of $X$  depends on $\bm{P}$, we will abuse $ X (\bm{P})$ and $X$ when they are clear from the context. We will also abuse $ L_{rand} (\bm{P}) $ and $L_{rand} $. 
 
Based on \eqref{eq: I-group-delay} and \eqref{eq: I-group-delay-e2}, we can calculate $ P(X\leq \ell) $ as follows,
\begin{equation} \label{eq: I-group-delay-e3}
P(X\leq \ell)=\prod_{i=1,...,K}P(Y_{i}(\bm{p}_i)\leq \ell).
\end{equation}

For $Y_i(\bm{p}_i)$, we have found the following result from the literature.
\begin{lemma} \label{lemma:coupon1}
\cite{anceaume2015new} 
%Let $\bm{p}=[p_0 \ p_1 \  \ldots \ p_{K-1}]$, where $p_i$ denotes the probability that coupon $i$ is drawn at a time slot, for $i=1,2,\ldots,K-1$, and $p_0$ denotes the probability that the null coupon is drawn. Let $Y$ denote the time needed to get a collection of $(K-1)$ coupons. Since the distribution of $Y$ depends on $\bm{p}$, we will use the notation $Y(\bm{p})$ instead of $ Y $. 
For any given $i$ and $ \ell\geq 0 $, if $ p(\mathcal{E}_i^j) =(1-p(\mathcal{E}_i^0))/(K-1)$ for $ j=1,\ldots,K-1 $, then
\begin{align*} 
&P(Y_i(\bm{p}_i)\leq \ell)= \notag \\
&1-\sum_{i=0}^{K-2}(-1)^{K-2-i}\binom{K-1}{i}\left[\dfrac{(K-1-i)p(\mathcal{E}_i^0)+i}{K-1} \right]^{\ell} .
\end{align*}
\end{lemma}

Next we  analyze $ L_{rand} $ under the optimized random scheme.  
%We will compare $ L_{rand} $ with the period $L$ of schedule sequences obtained by Construction $*$, for given $K$ and $M$. 
We have obtained in Section~\ref{sec: B-optimal random} that in the optimized random scheme,  for any node, the probability that another node successfully transmits to it in a time slot equals $P^*$ (see \eqref{eq: B-optimal-random}), that is, 
\begin{align*}
p(\mathcal{E}_i^1)=p(\mathcal{E}_i^2)=\cdots=p(\mathcal{E}_i^{K-1})=P^*, \\ p(\mathcal{E}_i^0)=1-(K-1)P^*,
\end{align*}
for any $i\in [K]$. Let $\bm{P}^*=[\bm{p}_1^* \ \bm{p}_2^* \ \ldots \ \bm{p}_K^*]$, where $ \bm{p}_i^*=[p(\mathcal{E}_i^0) \ \underbrace{P^*\ \ldots \  P^*}_{K-1}] $, for any $i\in [K]$. 
Then by  Lemma~\ref{lemma:coupon1} and~\eqref{eq: I-group-delay-e3}, we  obtain that for any $\ell \geq 0$,
\begin{align} 
&P(X(\bm{P}^*)\leq \ell)= \notag \\ 
&\left[ 1-\sum_{i=0}^{K-2}(-1)^{K-2-i}\binom{K-1}{i}\left[\dfrac{(K-1-i)p(\mathcal{E}_i^0)+i}{K-1} \right]^{\ell} \right]^K. \label{eq:random-g-e1}
\end{align}
Based on \eqref{eq:random-g-e1}, we can find  $L_{rand}(\bm{P}^*)$ to satisfy $P(X(\bm{P}^*)\leq L_{rand}(\bm{P}^*))=99.999\%$. We have listed $L_{rand}(\bm{P}^*)$ for some $K$ and $M$ in Table~\ref{tab: rand-prob6}.

Note that in the sequence scheme, even though~$L$ obtained from Construction $*$ is asymptotically decreasing with respect to~$W$ when $W\leq M'$, there are some cases where~$L$ with a larger~$W$ is longer than that with a smaller~$W$ due to the irregularity in occurrence of prime numbers. Therefore, given $M$, we choose the smallest one among $L$'s corresponding to  $W=1,2,\ldots,M$. We take the case of $K=10$ and $M=2$ for example. By Construction $*$, when $W=1$, $L=209$; while when $W=2$, $L=308$. Then given $M=2$, we only use one channel and thus $L=209$. We have also listed $L$ for some $K$ and $M$ in Table~\ref{tab: rand-prob6}, in which we have shown $ P(X(\bm{P}^*)\leq L) $ as well.

\begin{table}[h]
\begin{center}
\caption{ %$P(X_{i,j}\leq L_{seq})$ for Assignment T random scheme with the same duty factor as the sequence scheme. 
$L$ under Construction $*$ and $ L_{rand}(\bm{P}^*) $ under the optimized random scheme.  }
\label{tab: rand-prob6}
\begin{tabular}{|c|c|c|c|}
\hline
$(K,M)$ & $ L $ & $L_{rand}(\bm{P}^*)$ & $P(X(\bm{P}^*)\leq L)$  \\
\hline
$(10,1)$ & 209 & 406 & 0.9769  \\
\hline
$(10,2)$ & 209 & 406 & 0.9769  \\
\hline
$(15,1)$ & 493 & 656 &  0.9993  \\
\hline
$(15,3)$ & 462 & 656 &   0.9985  \\
\hline
$(18,1)$ & 665 & 812 & 0.9998  \\
\hline
$(18,2)$ & 665 & 812 & 0.9998 \\
\hline 
$(18,3)$ & 546 & 812 & 0.9972 \\
\hline
$(20,1)$ & 897 & 917 & 0.99998 \\
\hline
$(20,4)$ & 616 & 917 & 0.997 \\
\hline
$(24,1)$ & 1363 & 1130 & 0.999999 \\
\hline
$(24,3)$ & 1122 &  1130 & 0.99998 \\
\hline
$(24,4)$ & 728 & 1130 & 0.9944 \\
\hline
\end{tabular}
\end{center}
\end{table}

%By comparing Table~\ref{tab: rand-prob6} with Table~\ref{tab: rand-prob5}, we can find that $L_{rand}(\bm{p}^*)$ under random schemes with optimized transmitting probabilities is much shorter than $L_{rand}(\bm{p}_a)$ and $L_{rand}(\bm{p}_b)$, which are obtained under random schemes with the same power consumption as the sequence scheme.
From~Table~\ref{tab: rand-prob6}, we can observe that in most cases, the frame length under our proposed sequence scheme is shorter than that under the optimized random scheme, that is, $L< L_{rand}(\bm{P}^*) $. 
%It is notable that in these cases, although $ L_{rand}(\bm{P}^*) $ is substantially larger than $L$, the value of $ P(X(\bm{P}^*)\leq L )$ is already close to $99.999\%$. 
%This implies that to further improve the value of $ P(X(\bm{P}^*)\leq L_{rand}(\bm{P}^*))$ to 1, requires increasing $ L_{rand}(\bm{P}^*) $ significantly.
There exist some cases where $L_{rand}(\bm{P}^*)  <L$. For example, when $K=24$, $M=1$, we have $L_{rand}(\bm{P}^*)=1130$, $L=1363$. 
However, we should note that $L_{rand}(\bm{P}^*)$ just indicates that $  P(X(\bm{P}^*)\leq L_{rand}(\bm{P}^*))=99.999\% $, but cannot provide a hard guarantee on broadcast delay due to its probabilistic nature. 
In this case, even if we set $L_{rand}=L=1363$, we only have $ P(X(\bm{P}^*)\leq L_{rand})=99.9999\% $, instead of $ P(X(\bm{P}^*)\leq L_{rand})=1$. 

We can conclude that in terms of frame length, our proposed sequence scheme outperforms the random scheme in two aspects. One is that the sequence scheme can efficiently utilize multi-channel resources to reduce frame length while the random scheme cannot. The other is that the sequence scheme can provide a hard guarantee on delay.
\subsection{Broadcast completion time}
In this section, we consider another performance metric -- broadcast completion time. In order to show the relationship between the broadcast completion time and the number of employed channels, $W$, we let $W=1,2,\ldots,M$ in each scheme. For the general random scheme, we find $p_a$ to optimize $ P_a $ in \eqref{eq:random-e4}. For the Assignment~T based random scheme, we assume $P_{\alpha}=P_{\beta}$ and find  $p_b$ to optimize $P_{\beta}$ in \eqref{eq:AT-rand-e5}. Since we observe that the optimized $P_{\beta}$ is no less than the optimized $ P_a $ for any given $K$ and $W$, we will only compare the Assignment~T based random scheme with the sequence scheme. 

%\vspace{-4mm}
\begin{figure}[htbp]
    \centering
    \includegraphics[width=3.8in]{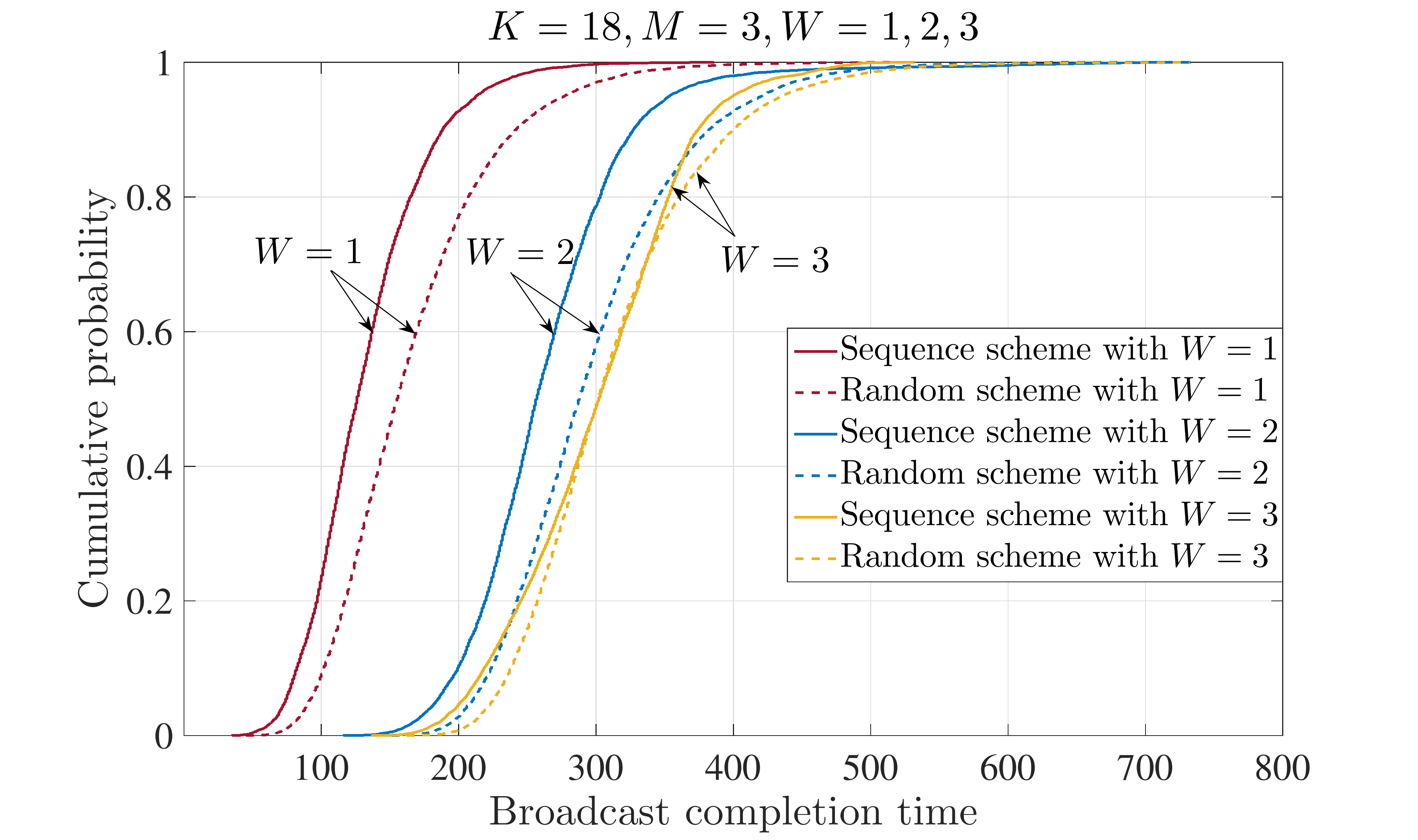}
  \caption{Broadcast completion time under sequence scheme and random scheme.}
\label{fig:broadcast completion time}
\end{figure}
%\vspace{-5mm}

Fig.~\ref{fig:broadcast completion time} shows the probability distribution of the broadcast completion time in 10000 runs for the case where $ K=18 $, $M=3$ and $W=1,2,3$ under the sequence scheme and the Assignment T based random scheme. The time offset of each node in each run is randomly generated. We can observe from Fig.~\ref{fig:broadcast completion time} that for both sequence scheme and random scheme, using only one channel can achieve shorter broadcast completion time with higher probability. We have conducted simulations for many other cases and observed the same result. For the random scheme, this is not surprising since we have obtained similar result  when we discuss frame length. However, for the sequence scheme, this is an interesting phenomenon and is contradictory to the performance for unicast completion time we considered for unicast in \cite{liu2019sequence}. In \cite{liu2019sequence}, we have shown that the sequence scheme can utilize multi-channel resources to decrease sequence period as well as the unicast completion time. The reasons behind may lie in the nature of broadcast and unicast and the tradeoff caused by multiple channels. We will try to explore the cause of this in the future. 

\section{Conclusion} \label{section: conclusion}
We investigate schedule sequence design to guarantee successful broadcast in an asynchronous ad hoc network. Previous works on the sequence design for broadcast are mainly developed with a single channel. In this paper, we derive a lower bound on the shortest common period and propose a CRT-based sequence construction method, for the multi-channel model. 
Under even group division with $W=M\leq M'$,
%When the number of channels is small, 
the period under our proposed construction has the same order as the lower bound.  We also achieve an asymptotic reduction in the order of $M$ compared with 
the shortest known sequence period for the single channel case.

We also analyze optimal transmitting and receiving probabilities for two random schemes. Comparisons for frame length and broadcast completion time under different schemes are conducted.  By comparison, we find that our proposed sequence scheme can ensure successful broadcast within shorter frame length than the optimized random scheme. Moreover, our proposed sequence scheme can decrease the frame length by utilizing multiple channels while the random schemes cannot. However, %the completion time performance is different from that for unicast.
using more channels would result in longer broadcast completion time, for  both sequence scheme and random scheme. 
%The comparison results can provide guidelines for our proposed sequence scheme. For systems that lay more emphasis on completion time, for example, systems where feedback information is available such that a node can start transmitting the $(i+1)$-th packet immediately after it knows that the $i$-th packet has been successfully delivered to its neighbors, then the sequence scheme for single channel is preferred since it can provide shorter completion time. While for systems that focus on frame length, then our proposed multi-channel sequence scheme is preferred since it can provide shorter frame length than the single channel case. 

%\begin{comment}
\appendix

\section{Proof for Lemma~\ref{thm: P-blocking}}
\label{appendix 1}
\begin{IEEEproof}
The sequence $(b_r)_{r=1}^{\infty}$ is non-negative and monotonically non-increasing. The difference between two adjacent entries in  $(b_r)_{r=1}^{\infty}$  is also monotonically non-increasing.
We let $ \lambda=\left \lceil \mu b_{1}^2/L \right \rceil $, which is the largest difference between two adjacent entries in the sequence $ (b_{r})^{\infty}_{r=2} $, and for $j= 1, 2, \ldots, \lambda$, let $n_i$ be the number of indices $r\geq 1$ such that $ b_r - b_{r+1}=i $. We have the following identity
\begin{equation}
n_{1}+2n_{2}+\cdots+\lambda n_{\lambda}=b_1.
\end{equation}
%We note that the value of $n_j$ may be equal to zero. 
We denote the largest $ b_{r} $ in $ (b_{r})^{\infty}_{r=1} $ such that $ b_{r}-b_{r+1}=i $ by $ b_{r_{i}^{+}} $, and denote the smallest such $ b_{r} $ by $ b_{r_{i}^{-}} $. The two entries followed by $ b_{r_{i}^{-}} $ are $ b_{r_{i}^{-}+1} $ and $ b_{r_{i}^{-}+2} $. For $ b_{r_{i}^{+}} $, we have
\begin{gather*}
b_{r_{i}^{+}}=b_1-\sum_{j=i+1}^{\lambda}jn_{j}.
\end{gather*}
For $ b_{r_{i}^{-}+1} $ and $ b_{r_{i}^{-}+2} $, we have
\begin{gather}
b_{r_{i}^{-}+1}-b_{r_{i}^{-}+2}=\left\lceil \dfrac{\mu b_1 b_{r_{i}^{-}+1}}{L} \right\rceil \leq i-1. \label{appendix: P-e0}
\end{gather}
The inequality in \eqref{appendix: P-e0} indicates
$$
b_{r_{i}^{-}+1} \leq \dfrac{(i-1)L}{\mu b_1}. 
$$
Since $  b_{r_{i}^{+}}-b_{r_{i}^{-}+1}=in_{i}$, then we have
$$ 
in_{i} \geq \left(b_1-\sum_{j=i+1}^{\lambda}jn_{j}\right)-\dfrac{(i-1)L}{\mu b_1},
$$ 
that is,
\begin{equation}\label{appendix: P-e1}
\dfrac{(i-1)L}{\mu b_1}+\sum_{j=i}^{\lambda}jn_{j} \geq b_1.
\end{equation}
For $ i=2,3,...,\lambda$, by dividing both sides of \eqref{appendix: P-e1} by $ i(i-1) $, and summing up the resulting inequalities, we have
\begin{equation} \label{appendix: P-e2}
\sum_{i=2}^{\lambda}\dfrac{L}{ \mu b_1 i}+\sum_{i=2}^{\lambda}\sum_{j=i}^{\lambda}\dfrac{jn_{j}}{i(i-1)}\geq \sum^{\lambda}_{i=2}\dfrac{b_1}{i(i-1)}.
\end{equation}
The RHS of \eqref{appendix: P-e2} is equal to
$$
\sum^{\lambda}_{i=2}\dfrac{b_1}{i(i-1)}=b_1 \left(1-\dfrac{1}{\lambda} \right),
$$
and the double summation in \eqref{appendix: P-e2} is equal to
$$
\sum_{i=2}^{\lambda}\sum_{j=i}^{\lambda}\dfrac{jn_{j}}{i(i-1)} = \sum_{j=2}^{\lambda}jn_{j} \sum_{i=2}^{j}\dfrac{1}{i(i-1)} =\sum_{j=2}^{\lambda} n_j(j-1). 
$$
Therefore we can rewrite \eqref{appendix: P-e2} as
\begin{gather*}
\dfrac{L}{\mu b_1}\sum^{\lambda}_{i=2}\dfrac{1}{i}+\sum_{j=2}^{\lambda}n_{j}(j-1) \geq b_1\left(1-\dfrac{1}{\lambda}\right),\\
\sum_{j=1}^{\lambda}n_{j}\leq \dfrac{b_1}{\lambda}+\dfrac{L}{\mu b_1}\sum^{\lambda}_{i=2}\dfrac{1}{i}.
\end{gather*}

If $ b_{C}\geq 1$, then the number of strictly positive differences between two adjacent entries in $ (b_{r})^{\infty}_{r=1} $ must be no less than $ C $, that is, $ C \leq \sum^{\lambda}_{j=1} n_{j}$. Thus, we have
\begin{equation}\label{appendix: P-e3}
C \leq \dfrac{b_1}{\lambda}+\dfrac{L}{\mu b_1}\sum_{i=2}^{\lambda}\dfrac{1}{i}.
\end{equation}
Note that when $\lambda=1$, \eqref{appendix: P-e3}  still holds since it is reduced to $ C\leq b_1 $. 

The inequality in \eqref{appendix: P-e3} can be re-written as
\begin{equation} \label{appendix: P-e4}
C\leq \sqrt{\dfrac{L}{\mu}}\left(\dfrac{b_1}{\lambda\sqrt{\dfrac{L}{\mu}}}+\dfrac{\sqrt{\dfrac{L}{\mu}}}{b_1} \sum_{i=2}^{\lambda}\dfrac{1}{i} \right).
\end{equation}
Let $ z=b_1/\sqrt{\dfrac{L}{\mu}} $. Then we write \eqref{appendix: P-e4} as 
$$
C\leq \sqrt{\dfrac{L}{\mu}} \left(\dfrac{z}{\lambda}+\dfrac{1}{z}\sum_{i=2}^{\lambda}\dfrac{1}{i} \right),
$$
 where  $ \lambda=\lceil z^2 \rceil $, that is, $ \sqrt{\lambda-1}<z\leq \sqrt{\lambda} $. 
Now we partition $\mathbb{R}_{+}$ into subintervals $I_{d}=(\sqrt{d-1},\sqrt{d}]$ for $ d=1,2,3,\ldots $, and let $ F:\mathbb{R}_{+}\rightarrow\mathbb{R}_{+} $ be a piecewise function defined as 
$$ F(x)=\dfrac{x}{d}+\dfrac{1}{x}\sum^{d}_{i=2}\dfrac{1}{i}, \text{ for } x\in I_{d}, d=1,2,3,\ldots . $$
As shown in Figure~\ref{fig:appendix P-2}, the function $ F(x) $ attains global maximum at $ x=\sqrt{2} $, with maximal value $ F(\sqrt{2})=3/\sqrt{8} $.  Thus 
$$
C \leq \sqrt{\dfrac{L}{\mu}}F(x) \leq\dfrac{3}{\sqrt{8}}\sqrt{\dfrac{L}{\mu}}.
$$
Therefore we can obtain that $ L \geq \left \lceil \dfrac{8C^{2}\mu}{9} \right\rceil $.

\begin{figure}[!h]
\centering
\includegraphics[width=2.5in]{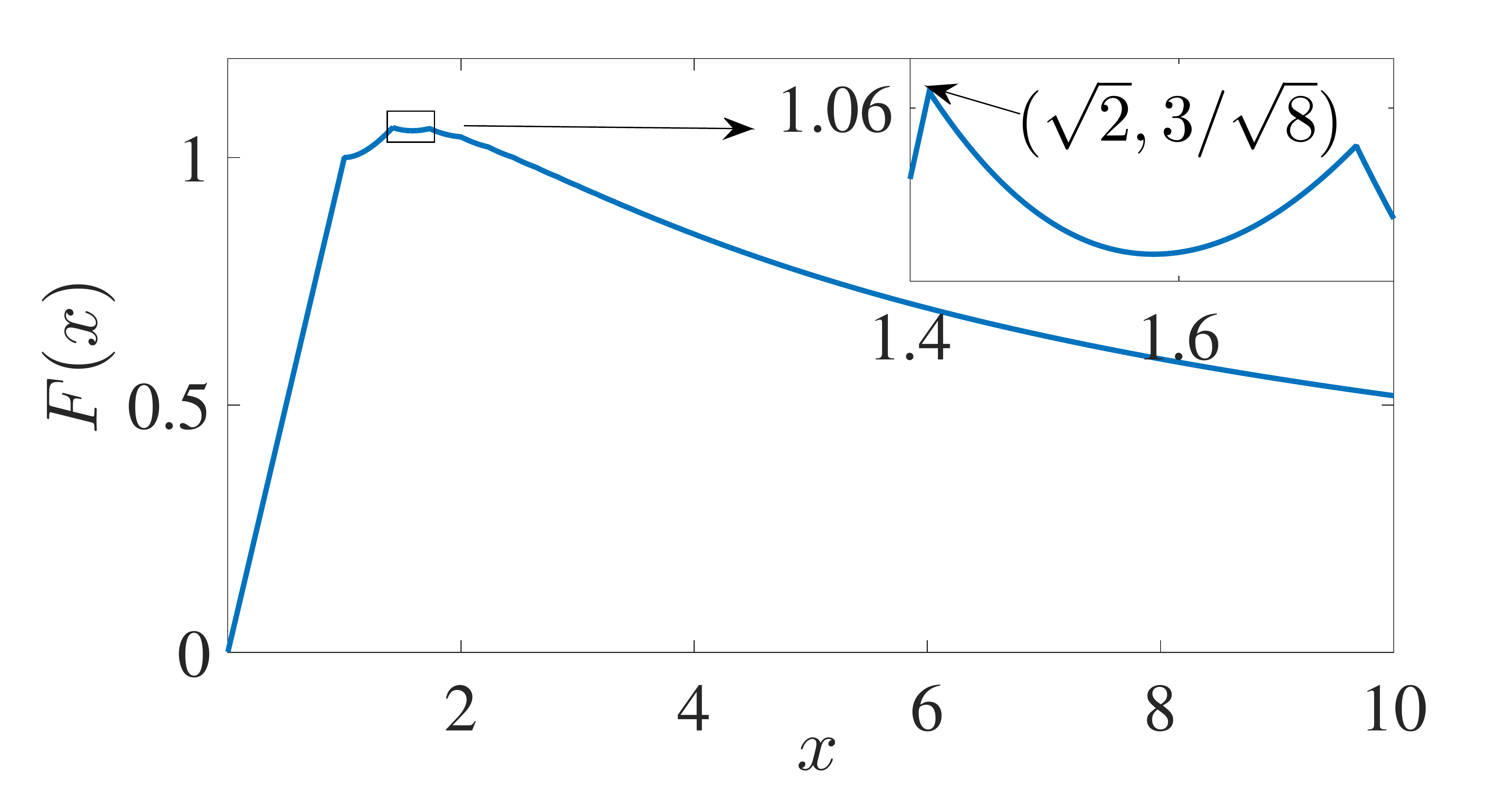}
\caption{The image of $F(x)$.}
\label{fig:appendix P-2}
\end{figure}
\end{IEEEproof}

%\end{comment}

%\begin{acknowledgements}
%\end{acknowledgements}


\vspace{-0.5cm}
\bibliographystyle{IEEEtran}
\bibliography{ref,ref-broadcast,ref-new}

\end{document}